\RequirePackage{rotating}
\documentclass[manuscript, screen]{acmart}

\usepackage{booktabs} 
\usepackage[ruled]{algorithm2e} 
\usepackage{subcaption}
\usepackage[printwatermark]{xwatermark}
\usepackage{xcolor}
\usepackage{graphicx}
\usepackage{tikz}
\usepackage{nameref}

\newsavebox\mybox
\savebox\mybox{\tikz[color=red,opacity=0.3]\node{WORKING DRAFT};}

\acmJournal{CSUR}
\acmVolume{01}
\acmNumber{01}
\acmArticle{01}
\acmYear{2017}
\acmMonth{07}

\setcopyright{rightsretained}

\acmDOI{0000001.0000001}

\begin{document}
\title{``I can assure you [\ldots] that it's going to be all right''} 
 \titlenote{HAL 9000, 2001 A Space Odyssey, full quote: ``I know everything hasn't been quite right with me, but I can assure you now, very confidently, that it's going to be all right again.
 }
 \subtitle{A definition, case for, and survey of algorithmic assurances in human-autonomy trust relationships}
\author{Brett Israelsen}
    \authornote{The corresponding author}
    \orcid{0000-0003-1602-1685}
    \email{brett.israelsen@colorado.edu}
    \affiliation{%
        \institution{University of Colorado, Boulder}
        \institution{Department of Computer Science}
        \city{Boulder}
        \country{USA}
    }
    \affiliation{%
        \institution{RECUV}
    }
    \affiliation{%
        \institution{C-UAS}
    }

\begin{abstract}
    As technology become more advanced, those who design, use and are otherwise affected by it want to know that it will perform correctly, and understand why it does what it does, and how to use it appropriately. In essence they want to be able to \emph{trust} the systems that are being designed. In this survey we present \emph{assurances} that are the method by which users can understand how to trust this technology. Trust between humans and autonomy is reviewed, and the implications for the design of assurances are highlighted. A survey of existing research regarding assurances is presented, and several key ideas are extracted in order to refine the definition of assurances. Several directions for future research are identified and discussed.
\end{abstract}

%
%

%
%



\thanks{This work is supported by C-UAS and Northrop-Grumman Aerospace Systems.}


\maketitle

\section{Introduction}
    As technology becomes more advanced, those who design, use, and are affected by it in other ways want to know that it will perform correctly, and understand why it does what is does, and how to use it appropriately. In essence, people who interact with advanced technology want to be able to trust it appropriately, and then act on that trust.

    In interpersonal relationships, and otherwise, humans act largely based on trust. For example, a supervisor asks a subordinate to accomplish a task based on several factors that indicate they can trust them to accomplish that task. When consumers make purchases, they do so with trust that the product will perform as promised. Likewise, when using something like an autonomous vehicle, the user must be able to trust it appropriately in order to use it properly.

    With the rapid advancement of the capabilities of intelligent computing technology to do tasks that were previously assumed to be too complicated for computers, there has been much recent discussion regarding how humans can trust this technology -- although the connection to trust is not always made explicit, per se. This discussion has taken place both in public \cite{Spectrum2016-jv,DeSteno2014-cq,Cranz2017-yh,Cassel2017-tn,Danks2017-sb,Wagner2016-ck}, business \cite{Banavar2016-nm, Khosravi2016-ke,Moody2017-vd,Rudnitsky2017-in,Benioff2016-tc,Tankard2016-rk}, and academic \cite{Groom2007-bz,Lloyd2014-bb,Goodrum_2016-fm,Foley2017-qj,Ghahramani2015-yq,Castelvecchi2016-mr} settings.

    Those who discuss \emph{how} to trust a specific technology are really referring to the need for some indication of the appropriate level of trust to give said technology. In other words, it is desirable to \emph{design} capabilities and methods for intelligent technology which help us achieve appropriate levels of trust in that technology. These capabilities and methods are collectively referred to as \emph{assurances}.
    
    Specifically, this survey investigates what assurances an Artificially Intelligent Agent (AIA) can provide to a human user in order to affect their trust. The colloquial definitions of `appropriate use', `assurance', `AIA', and `trust' should suffice for now to give the reader a general idea of the motivation; more formal definitions will be presented in Section \ref{sec:background}. It is the author's position that there are many researchers, from different disciplines, who will potentially be interested in this work. This group includes those who are interested in working with, trusting, interpreting, understanding, and/or regulating AIAs.

\begin{figure}
        \centering
        \includegraphics[width=0.7\textwidth]{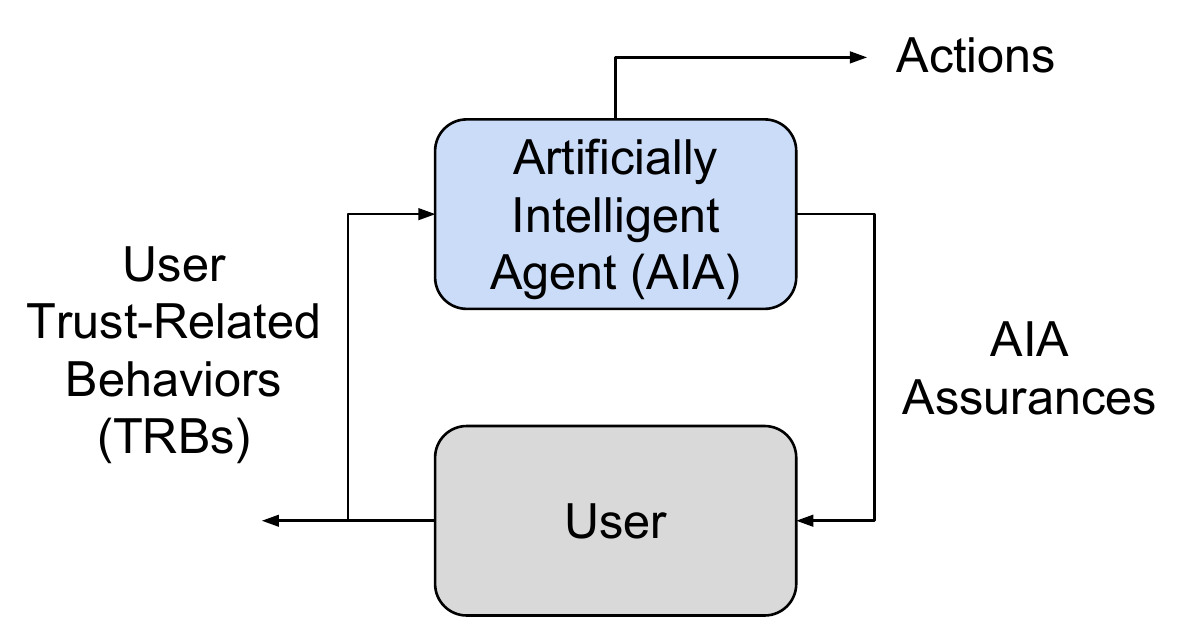}
        \caption{Diagram depicting the simple one-way trust development relationship between a human user and an AIA. Based on a user's level of trust they take certain actions (e.g. give AIA commands), these commands can lead the AIA to certain actions and/or to provide assurances to the user in order to affect their trust.}
        \label{fig:SimpleTrust_one_way}
    \end{figure}

    Figure~\ref{fig:SimpleTrust_one_way} is a simple diagram of the trust cycle that exists between a human user and an AIA (justification for the existence of this cycle will be presented later). Simply, the user's trust is affected by assurances that in turn affect the user's behaviors in interacting with the AIA (e.g. to trust AIA with responsibilities, or not). To fully understand and appreciate the importance of assurances, one must have a more formal understanding of each component in Figure~\ref{fig:SimpleTrust_one_way}.

    This paper provides an overview of the components of Figure~\ref{fig:SimpleTrust_one_way}. It then turns a more focused attention to assurances, and investigates some of the related research that has been done to date. From this survey of literature, properties and classifications of assurances are created, and directions and considerations for further research are presented.

    Some of the novel contributions of this paper include: identifying several different bodies of research that contribute to this topic; a detailed description and definition of assurances in general human-AIA relationships; and a detailed breakdown of the different components of assurances, and design considerations for implementing them. To this end, Section~\ref{sec:background} provides definitions for each of the terms. In Section~\ref{sec:methodology} we discuss the methodology used when compiling this survey. Afterwards, Section~\ref{sec:survey} will discuss the current landscape of assurances that exist in the literature. Section \ref{sec:synthesis} draws important insights and conclusions from the literature, and then uses those to present a more detailed version of Figure~\ref{fig:SimpleTrust_one_way}, that helps inform considerations when classifying and designing assurances. Finally, conclusions are presented in Section~\ref{sec:conclusions}.

\section{Background and Motivation} \label{sec:background}

What do people who talk about `comprehensible systems', `interpretable learning', `opaque systems', and `explainable AI' really care about? This document investigates, in a formal and comprehensive manner, which dimensions of a user's trust should be influenced by the assurances of an AIA, as well as how this might be practically accomplished.

Motivations and a grounding application are presented is Sections~\ref{sec:motivation} and \ref{sec:mot_example}, along with some application-specific definitions of the trust-cycle terms. A discussion of related work is in Section~\ref{sec:rel_work}. Assurances are presented and defined in Section~\ref{sec:assurances}.  Section \ref{sec:aias} defines the term AIA in a general sense, and how they relate to assurances. Section \ref{sec:trust} discusses trust, and provides a model of human-AIA trust that can be used in designing assurances. Finally, Section~\ref{sec:trbs} discusses trust-related behaviors and how they should be considered when designing assurances.

\subsection{Motivation} \label{sec:motivation}
    Generally, humans have always wanted to trust the tools and systems that they create.  To this end many metrics and methods have been created to help assure the designers and users that the tools and systems are in fact capable of being applied in certain ways, and/or will behave as expected. Of course, at this point in the document `trust' is quite an imprecise term, and needs to be more formally defined; this will be done in Section~\ref{sec:trust}

    The situation has grown more complicated in recent years because the advanced capabilities of the systems being created can at times be difficult for even those who designed them to comprehend and predict. There are, for instance, systems that have been designed to learn from extremely large amounts of data and expected to regularly perform on never before seen data. In some cases, such systems have been designed to perform tasks that might take humans entire lifetimes to complete. Below is a small sample of some application areas that exist and a possible reason why they -- perhaps unknowingly -- have an interest in creating trustworthy systems.

    \begin{description}
        \item [Interpretable Science:] Scientists need to be able to trust that the models created using data analysis, and be able to draw insights from them. Scientific discoveries cannot depend on methods that are not understood.
        \item [Reliable Machine Learning:] It is critical to have safety guarantees for AIAs that have been deployed in real-world environments. Failing to do so can result in serious accidents that could cause loss of life, or significant damage.
        \item [Artificial Intelligence/Machine Learning:] There is a need to interpret how and why theoretical AIA models function. This is due to the need to know they are being applied correctly, but also to be able to design new methods to overcome weaknesses in the existing methods.
        \item [Government:] Governments are beginning to enforce regulations on the interpretability of certain algorithms in order to ensure that citizens can understand why many services make the decisions and predictions that they do. A specific example are the algorithms deployed by credit agencies to approve/reject loans.
        \item [Medicine:] Medical professionals need to understand why data-driven models give predictions so that they can choose whether or not to follow the recommendations. While AIAs can be a very useful tool, ultimately doctors are liable for the decisions they make and treatments they administer.
        \item [Cognitive Assistance:] Systems are being designed as aids for humans to make complex decisions, such as searching and assimilating information from databases of legal proceedings. When an AIA presents perspectives and conclusions as summaries of this data, it must be able to also present evidence and logic to justify them.
    \end{description}

    The interests of the author lie specifically in the design of unmanned vehicles that operate in concert with human operators in uncertain environments. In this setting, it is desirable for the unmanned vehicle to be able to communicate with a human in some way in order to help them properly use the vehicle. The hope is that, in doing so, the performance of the team can be improved by appropriately utilizing the strengths of both the human and unmanned vehicle. This application is explained in more detail below in relation to Figure~\ref{fig:SimpleTrust_one_way}.

    Whether formally acknowledged or not humans want to design assurances to help them appropriately trust AIAs (a point already mentioned in the introduction). There are a few research fields that have formally and explicitly considered trust between humans and technology. Some examples are: e-commerce, automation, and human-robot interaction. However, due to their main goals these research efforts have mainly focused on implicit properties of the systems that affect trust. Conversely, there are other research fields that have informally considered how to affect the trust of designers and users via explicitly designed assurances. However, due to their informal treatment of trust, it is unknown and unclear how effective these designed assurances might be in practice, or what principles ought to be considered when designing assurances for general AIAs. A key objective of this paper is to survey these areas and, in doing so, help bridge the gap between them by identifying common goals and approaches, as well as highlighting where the different groups might benefit from the research of others.

\subsection{Motivating Application and Basic Definitions} \label{sec:mot_example}
    It is useful to have a concrete example on which we can refer for grounding examples. As previously mentioned the specific interests of the author lie in the design of unmanned vehicles that can work in cooperation with human operators.

    Specifically, consider an unmanned ground vehicle (UGV) in a road network with unmanned ground sensors (UGSs). The road network also contains a pursuer that the UGV is trying to evade while exiting the road network. A human operator monitors and interfaces with the UGV during operation. The operator does not have a detailed knowledge of how the UGV functions or makes decisions, but can interrogate the system, modify the decision making stance (such as `aggressive' or `conservative'), and provide information and advice to the UGV. In this situation the operator could benefit from the UGVs ability to express confidence in its ability to escape given the current sensor information, and work with the AIA to modify behavior if necessary. This application serves as a useful analog for many different autonomous system applications, e.g. autonomous VIP escort problems and intelligence, surveillance and reconnaissance (ISR)/counter-ISR operations for defense and security \cite{Kingston2012-va}.
   
	\begin{figure}[htbp]
    	\centering
     	\includegraphics[width=0.4\textwidth]{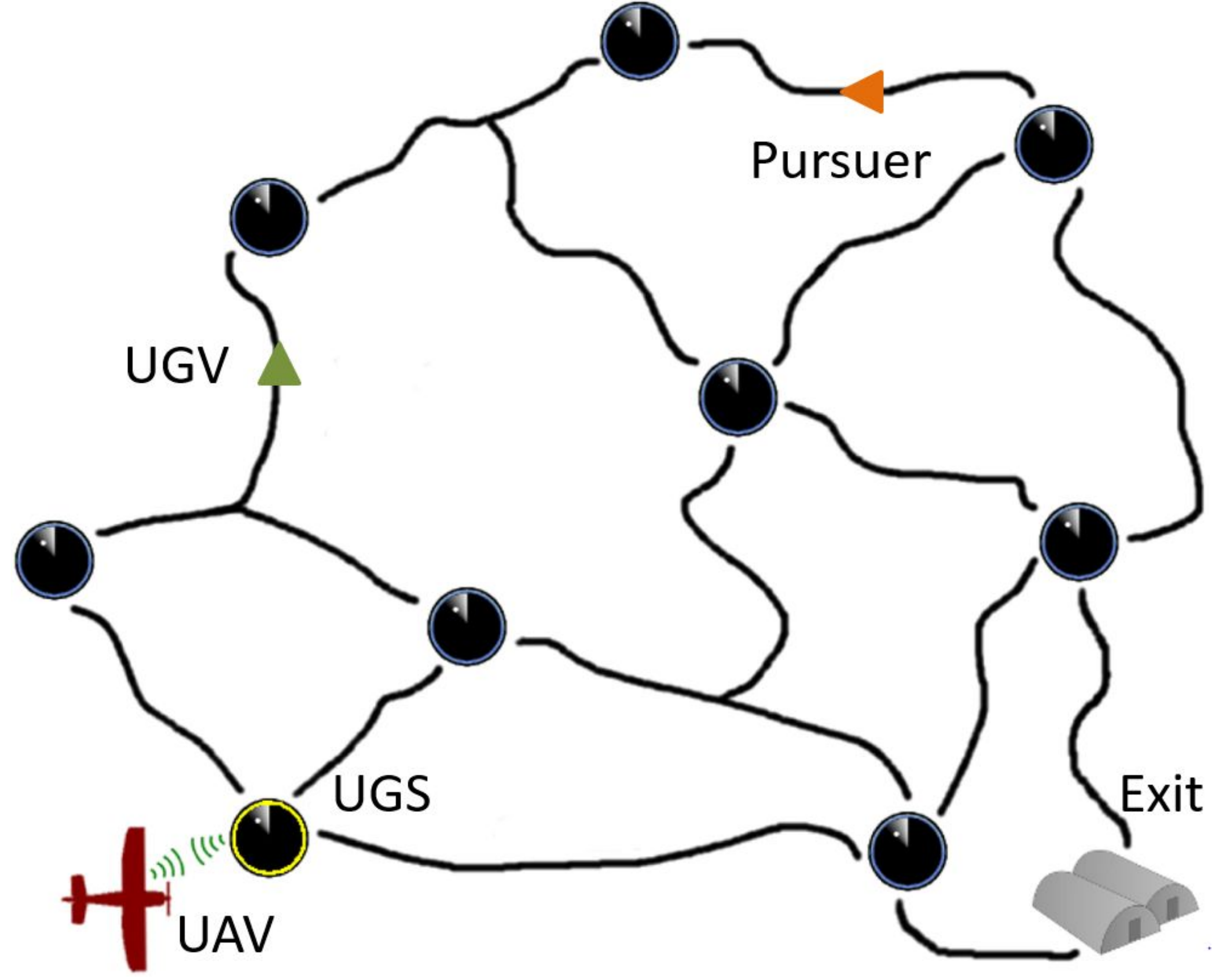}
    	\caption{Application example of unmanned ground vehicle (UGV) in a road network, trying to evade a pursuer. The UGV has access to unmanned ground sensors (UGSs) (also an unmanned aerial vehicle (UAV) that can be used as a mobile sensor), as well as information and decision making advice from a non-expert human operator. The operator's actions towards the UGV are trust-based.}
        \label{fig:RoadNet}
    \end{figure}

    In this scenario the trust-cycle terms can be simply defined as follows:

    \begin{description}
        \item [Artificially Intelligent Agent:] The UGV, this problem is difficult because the pursuer is only observed sporadically, and does not follow a known course. The UGV must make decisions under uncertainty, with little information. 
        \item [Trust:] The operator's willingness to rely on the UGV when the outcome of the mission is at stake, such as in a scenario where the UGV is carrying a valuable payload.
        \item [Trust-Related Behaviors:] The operator's behaviors towards the UGV, including the information provided, and the commands given. This might take the form of approving/rejecting the UGV plan upfront, or the possibility of involving real-time communication and adjustments to new information that the operator receives.
        \item [Assurances:] Implicit and explicit communication from the UGV that has an effect on the operator's trust. An assurance might be in the form of communicating the probability of success for a given plan, or communicating that the mission is not proceeding as expected.
    \end{description}

\subsection{Related Work}\label{sec:rel_work}
    There has been a fair amount of work recently in the AI community that considers the importance of interpretability, explainability, and transparency (i.e. \citet{Doshi-Velez2017-xy}, \citet{Weller2017-zx}, and \citet{Lipton2016-ug}, and \citet{Gunning2017-ih}). This work has many interesting and important insights regarding the need for transparency, but does not formally acknowledge the role of trust in human decision making or how interpretability and transparency affect the trust of those who use AIAs. However, this work is extremely beneficial because it draws the attention of researchers to this critical area, and it is the beginning of the formalization of the problem from those who are currently doing a majority of the design of assurances.

    \citet{Lillard2016-yg} addressed the role of assurances in the relationship between humans and AIAs, and provides much of the foundation for describing the relationships between assurances and trust in human-AIA interaction. Here, the framework for analyzing assurances is presented in a way that is both more general and more detailed, albeit with the same end goal of being applied in a very similar end application. For instance, we consider the full trust model presented by \citet{McKnight2001-fa}, whereas \citeauthor{Lillard2016-yg} only consider a subset of the trust model.

    Regarding the relationship with the work of \citet{McKnight2001-fa}, who constructed a typology of interpersonal trust, we adopt the position that besides being applied to the e-commerce industry their trust model also applies to relationships between humans and AIAs (as in \citet{Lillard2016-yg}). They refer to something called `vendor interventions' that are related to what we call assurances in this paper. One small, but important distinction between vendor interventions and assurances is that assurances cannot directly have an affect on the user trust-related behaviors (TRBs). The premise is that no effect on a user's trust can force a TRB. This is an important point because we are considering a scenario (arguably more realistic, and definitely more humane) in which the human and AIA are working together, and not where the human is a puppet cont

    \citet{Corritore2003-gx} refer to a concept similar to assurances which they call `trust cues'. They review some of the literature that existed in the domain of e-commerce at that time. The wrap these trust cues into what they call `external factors' that affect the trust of a user of the internet. The way they present trust cues is as properties of websites, and the effects that those properties had on customers.
    
    While assurances are defined by \citeauthor{Lillard2016-yg}, and mentioned by \citeauthor{McKnight2001-fa}, and \citeauthor{Corritore2003-gx}, we investigate in detail how assurances fit within the trust cycle (from Figure~\ref{fig:SimpleTrust_one_way}), survey what methods of assurance have been and are currently being used, then present a refined definition and classification of assurances. In essence others have noted the existence of assurances, and we now directly consider the question: What, exactly, are assurances, and how can they be designed? To that end we perform a survey of literature that formally considers trust between humans and AIAs, as well as literature that informally investigates trust through topics like transparency, explainability, and interpretability, and begin to distill ideas for practically designing assurances in human-AIA trust relationships.

\subsection{Assurances} \label{sec:assurances}
    The term assurances was introduced in the previous section as the name by which feedback will be known in a human-AIA trust relationship. As assurances are the main topic of this paper, and have received little attention in trust literature, a more detailed definition and discussion is merited.

    \citet{McKnight2001-fa} allude to this kind of feedback in an e-commerce relationship as `Web Vendor Interventions' and mention some possible actions that might be used in that specific application. They go as far as making a diagram that indicates that these interventions could affect the `Trusting Beliefs', `Trusting Intentions', and `Trust-Related Behaviors' (see Figure~\ref{fig:UserTrust}) of an online human user. \citet{Corritore2003-gx} refer to assurances as `trust cues' that can influence how online users trust vendors in an e-commerce setting. \citet{Lee2004-pv} discuss `display characteristics', which are methods by which an autonomous systems can communicate information to an operator. The term assurances is perhaps earliest used in the context of human-AIA relationships by \citet{Sheridan1984-kx}. More recently, and formally, \citet{Lillard2016-yg} defined the term `assurance', we extend this definition here to be more general:    
    \begin{description}
        \item [Assurance:] A property or behavior of an AIA that affects a user's trust. This affect can increase \emph{or} decrease trust. 
    \end{description}

    Most researchers familiar with the fields of AI, ML, data science, and robotics will recognize terms like \emph{interpretable}, \emph{comprehensible}, \emph{transparent}, \emph{verified and validated} (V\&V), \emph{certified}, and \emph{explainable AI}, with respect to the models or performance of a designed system. A key claim of this paper is that \textbf{from a high level all of these approaches have the same aim: for a user to be able to trust an AIA to operate in a certain way, and based on that trust behave appropriately towards the AIA}. These methods are therefore means of assuring or calculating assurances to be given to a human user.

    Refined concepts about assurances are presented in Section~\ref{sec:synthesis} which elaborates on Figure~\ref{fig:SimpleTrust_one_way}. There, important concepts and considerations for designing assurances are reviewed. 

\subsection{Artificially Intelligent Agents} \label{sec:aias}
    As noted by \citet{Tripp2011-rx} technology spans a wide spectrum of capabilities. With regards to autonomous systems one might consider anything from a thermostat, to HAL 9000. While the main interest of the author is geared towards the capability of humans to trust `advanced' technology, for the purposes of this survey we will take a more holistic view and use the term Artificially Intelligent Agent (AIA) to encompass a broad range of technologies that can be considered automatic by some sense of the word. This is done in order to provide generally applicable definitions.

\begin{figure}[htbp]
    	\centering
     	\includegraphics[width=0.7\textwidth]{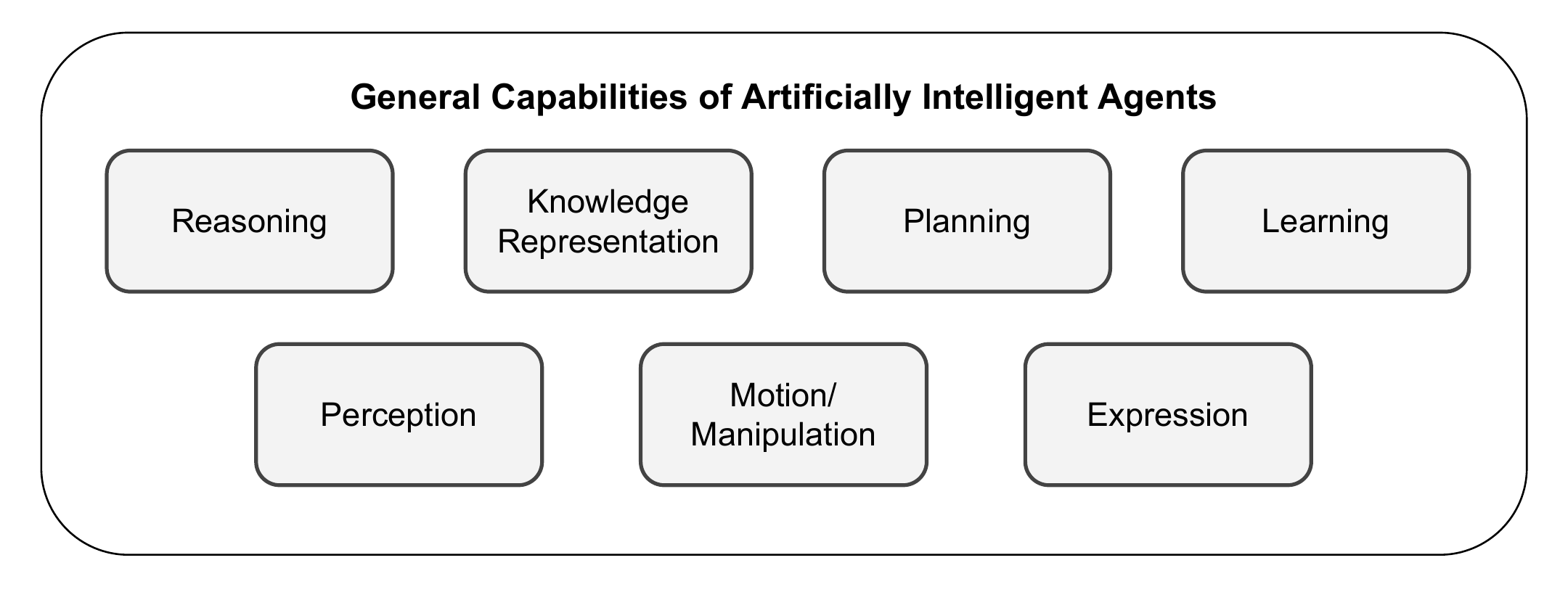}
    	\caption{List of the capabilities of an artificial general intelligence. In this paper an AI is defined as a system that possesses at least one of the capabilities illustrated. AIA capabilities are the sources of assurances.}
        \label{fig:AIcapabilities}
    \end{figure}

    It is generally accepted that an artificial intelligence needs to possess the capabilities shown in Figure~\ref{fig:AIcapabilities} \cite{Russell2010-wv,Nilsson2009-rp,Luger2008-vf}\footnote{This group of capabilities would generally be accepted in the AI community, although some argue that it is  necessary to add other categories like imagination, creativity, and social interaction. Of course this set of attributes is not universally accepted, and is still being refined.}. The following simple definitions will help to ground further discussion in the paper:

    \begin{description}
        \item [Reasoning:] The ability to solve problems, and make conclusions.
        \item [Knowledge Representation:] The ability to internally represent knowledge of information that has been learned.
        \item [Planning:] The ability to make a plan in order to accomplish a goal within an environment.
        \item [Learning:] The ability to learn from experience and data.
        \item [Perception:] The ability to use different sensors to perceive the surrounding environment.
        \item [Motion/Manipulation:] The ability to move within an environment and manipulate parts of it.
        \item [Interaction:] The ability to interact with other intelligent agents. For communicating with humans this could involve some type of natural language interface.
    \end{description}

    It should be noted that the categories are not clearly separable; for instance where does the capability to plan end, and reasoning begin? Similar questions could be asked of the other capabilities. Regardless, the concept of a separation is useful in defining an AIA:
    
    \begin{description}
        \item[Artificially Intelligent Agent:] an agent that acts on a goal (internally or externally generated), and possesses, to some extent, at least one of the capabilities shown in Figure~\ref{fig:AIcapabilities}.
    \end{description}

\begin{figure}[htbp]
    	\centering
     	\includegraphics[width=0.9\textwidth]{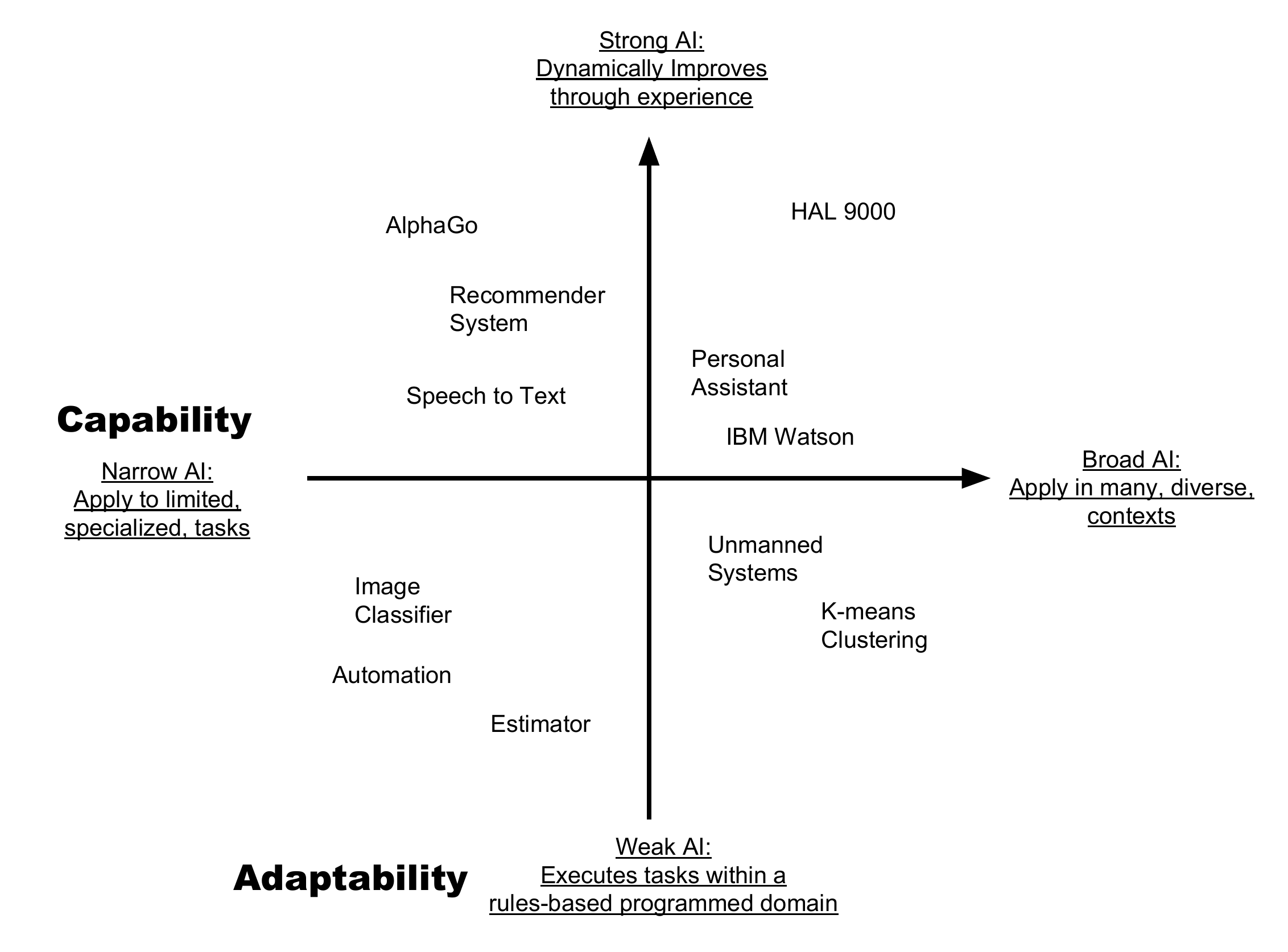}
    	\caption{Illustration of the range of systems encompassed by the AIA definition. Horizontal axis reflects the scope of the AIA, the vertical axis reflects the adaptability of the AIA.}
        \label{fig:StrongWeak}
    \end{figure}

    Of course, with such a broad definition, there will also be a broad range of different AIAs. It is argued here that it is most useful to classify that range along two axes: scope, and adaptability. The term scope refers to the range of possible applications of the AIA: does it have a certain specific application, or can it be used in many different applications? Agents with narrow scope are more specialized, while those with a broad scope can be applied in more diverse applications. Adaptability refers to the ability of the AIA to become better at executing its goal over time. Low adaptability has been termed `weak' and high adaptability `strong'. Figure~\ref{fig:StrongWeak} is a depiction of these two axes, and where some typical AIAs might fall in that space.

  Arguably we might just use the term artificial intelligence (AI) instead of AIA, however the term AI carries too much ambiguity (in its fullest meaning it would possess all capabilities from Figure~\ref{fig:AIcapabilities}, and more). Using AIA allows the broad inclusion of \emph{any} system in the adaptability/scope plane. Figure~\ref{fig:StrongWeak} shows some examples of AIAs and their places on the plane. To make sure the point is clear, the research discipline called machine learning (ML) is a subset of the AI research landscape. Individual ML algorithms might be thought of as being a narrowly scoped AI that is contained within only one of the AIA capabilities. As some examples, consider the following systems and the capabilities that they possess.

    \begin{enumerate}
         \item An autonomous bottle capping machine might \textbf{perceive} bottles, and \textbf{move} mechanical parts to place caps on them
         \item An unmanned aerial vehicle (UAV) might possess the ability to \textbf{plan} missions, \textbf{perceive} its location, and execute \textbf{motion} of its components to carry out a plan
         \item A virtual personal assistant might be capable of \textbf{interacting} with a user, \textbf{learning} the user's preferences, and \textbf{reasoning} about what assistance the user needs
         \item An image classifier might possess the capability to \textbf{learn} image classes from labeled examples and predict the class of never-before-seen new images.
     \end{enumerate}

    One might also question the need to define AIAs in the first place. This is to aid in the search for and understanding of assurances. As will be shown later, different methods of assurance can be found over the entire range of AIAs, so that an automation system such as a bottle capping machine might be able to use similar assurances -- or more generally, similar principles of assurance -- as might a self-driving car, and vice-versa. The capabilities of AIAs (shown in Figure~\ref{fig:AIcapabilities}) are the sources of assurances, in other words, assurances cannot exist without some source AIA capability.

    This definition, while broad, is still useful because it encompasses many of the systems that are typically described as `artificially intelligent'. More importantly, many of the assurances that exist for the simplest AIAs (e.g. a simple k-means classifier) can be extended for use in more advanced AIAs. In other words, the definition of AIAs sets an appropriate scope for the bodies of research that are likely to have investigated assurances and assurance principles that can be generalized/extended to any intelligent computing system. The definition of AIAs and their range of capabilities also helps to understand and establish what kinds of assurances might be needed in future systems. For example, assurances from an AIA that can only carry out planning tasks will probably differ in design and/or application from assurances from an AIA that can only carry out perception tasks. 

\subsection{User Trust} \label{sec:trust}
    In designing assurances that affect trust-based user behaviors, it is critical to know what drives those behaviors. Because of this, some time must be spent to understand what trust is. 

    Trust is critical in interpersonal relationships, and it affects the dynamics of intelligent multi-agent systems as simple as one-on-one personal interactions  \cite{Lewicki2006-hj}, to more complicated ones such as financial markets and governments \cite{Fukuyama1995-un}. Consequently, researchers in psychology, sociology, and economics have historically sought to understand the fundamental principles of trust, each with the aim of understanding their field better \cite{Gambetta1988-pi}. Moral philosophers have also thought intently about the topic \cite{Baier1986-im}.

    Due to wide interest spanning many disciplines it is difficult, if not impossible, to write a succinct definition of trust that would appease all interested parties. Besides that, trust is actually a very broad concept that evades precise definitions at a high level. However the following definition, adapted from \cite{McKnight2004-vv}, is broad enough to avoid too much contention:

    \begin{description}
        \item [Trust:] a psychological state in which an agent willingly and securely becomes vulnerable, or depends on, a trustee (e.g., another person, institution, or an AIA), having taken into consideration the characteristics (e.g., benevolence, integrity, competence) of the trustee.
    \end{description}

    \subsubsection{Trust between AIAs and humans?}
        Trust is generally understood to exist between people. Is it possible for a human to enter into a trusting relationship with an AIA? That humans actually do feel trust towards machines has been experimentally confirmed several times in research using common subjective psychological questionnaires. Some examples include: \citet{Muir1996-gt,Reeves1997-ad,Groom2007-bz,Mcknight2011-gv,Riley1996-qm,Bainbridge2011-pl,Kaniarasu2012-mo,Salem2015-md,Desai2012-rc, Freedy2007-sg, Inagaki1998-cl, Kaniarasu2013-ho}, and \citet{Wang2016-id}.

        Several academic experiments have investigated the possibility of trust existing between humans and (according to the terminology of this survey) AIAs. All found that some level of trust can be formed in such relationships. For instance, \citet{Lacher2014-yc} points out that people trust an AIA at different levels. As an example, an operator would have different perspectives on trust based on their level of interaction with the AIA. The designer of an AIA would also trust the AIA differently than an end user, due to the differing nature of the trust relationship from one to the other. 

        \citet{Tripp2011-rx} investigate the variation of trust between humans and different levels of technology. They run experiments with three different levels of technology: Microsoft Access, a recommender system, and Facebook. They found that `human-like' trust applied more to Facebook, while `system-like' trust applied more to MS Access. They conclude that if the system is `human enough', then a human trust model is appropriate. While there is technically a difference between their definitions of `human' and `system' trust, we argue that for practitioners the differences are negligible. As an example they suggest replacing the competence dimension of the human trust model with a dimension called functionality that has, in essence, the same definition as competence except it accounts for reduced complexity of the system.

        Therefore, in the interest of reducing complexity we will use a human trust model as a basis for human-AIA trust -- with the understanding that the definitions of the model must correspondingly vary with the complexity of the AIA.

\subsubsection{A Model of Human-AIA Trust}
        We now present a model of human trust, which will cast insights on the necessary effects of assurances. It should be noted that this model is being presented as \emph{one possible model} that can be helpful in understanding assurances -- it is neither the only model, or a perfect model (as discussed above). As research advances, such models will likely continue to evolve, and the ideas of assurances will naturally evolve as well.

\begin{figure}[htbp]
            \centering
            \includegraphics[width=0.9\textwidth]{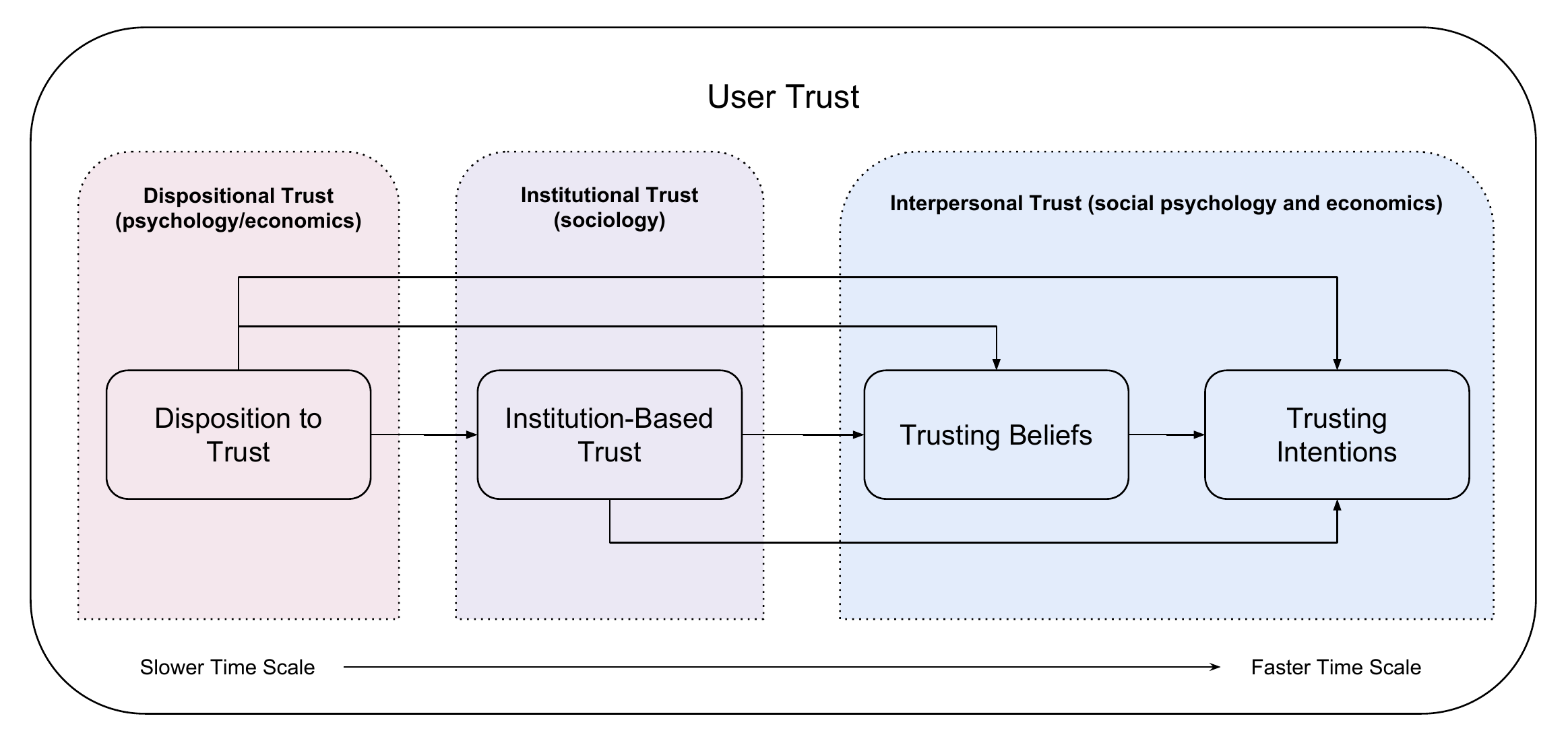}
            \caption{Interdisciplinary trust model proposed by \citet{McKnight2001-fa}. The three main categories are delineated, and corresponding disciplines that are interested are listed within parentheses. Connections indicate a causal relationship. The suggestion regarding time scales of the development of trust is the author's addition; trust development is discussed more in \cite{Lewicki2006-gp}, and \cite{Lewicki2006-hj}}
            \label{fig:UserTrust}
        \end{figure}

        In work relating to business management, \citet{McKnight1998-ty}, and later \citet{McKnight2001-fa}, performed what is, arguably, the first multi-disciplinary survey and unification of trust literature, which also condensed it into a single typology. The resulting model is shown, with some minor adaptations, in Figure~\ref{fig:UserTrust}. The figure illustrates the three categories that make up a human's trust. There are causal arrows that connect the different components. The `Dispositional Trust' block is generally considered by psychologists, and deals with long-term psychological traits that develop in a person from childhood. The `Institutional Trust' block is generally studied by sociologists, and represents the level to which a person trusts social/commercial structures. Finally, the `Interpersonal Trust' block is deals directly with one-on-one relationships and can generally fluctuate more quickly than the other two.

        In the context of AIAs the components of the three categories from Figure~\ref{fig:UserTrust} are defined as follows:

        \begin{description}
            \item [Disposition to Trust:] The extent to which one displays a consistent tendency to be willing to depend on AIAs in general across a broad spectrum of situations and persons
            \item [Institution-Based Trust:] One believes that regulations are in place that are conducive to situational success in an endeavor
            \item [Trusting Beliefs:] One believes that the AIA has one or more characteristics beneficial to oneself
            \item [Trusting Intentions:] One is willing to depend on, or intends to depend on, the AIA even though one cannot control its every action
        \end{description}

\begin{sidewaysfigure}[htbp]
            \includegraphics[width=8in]{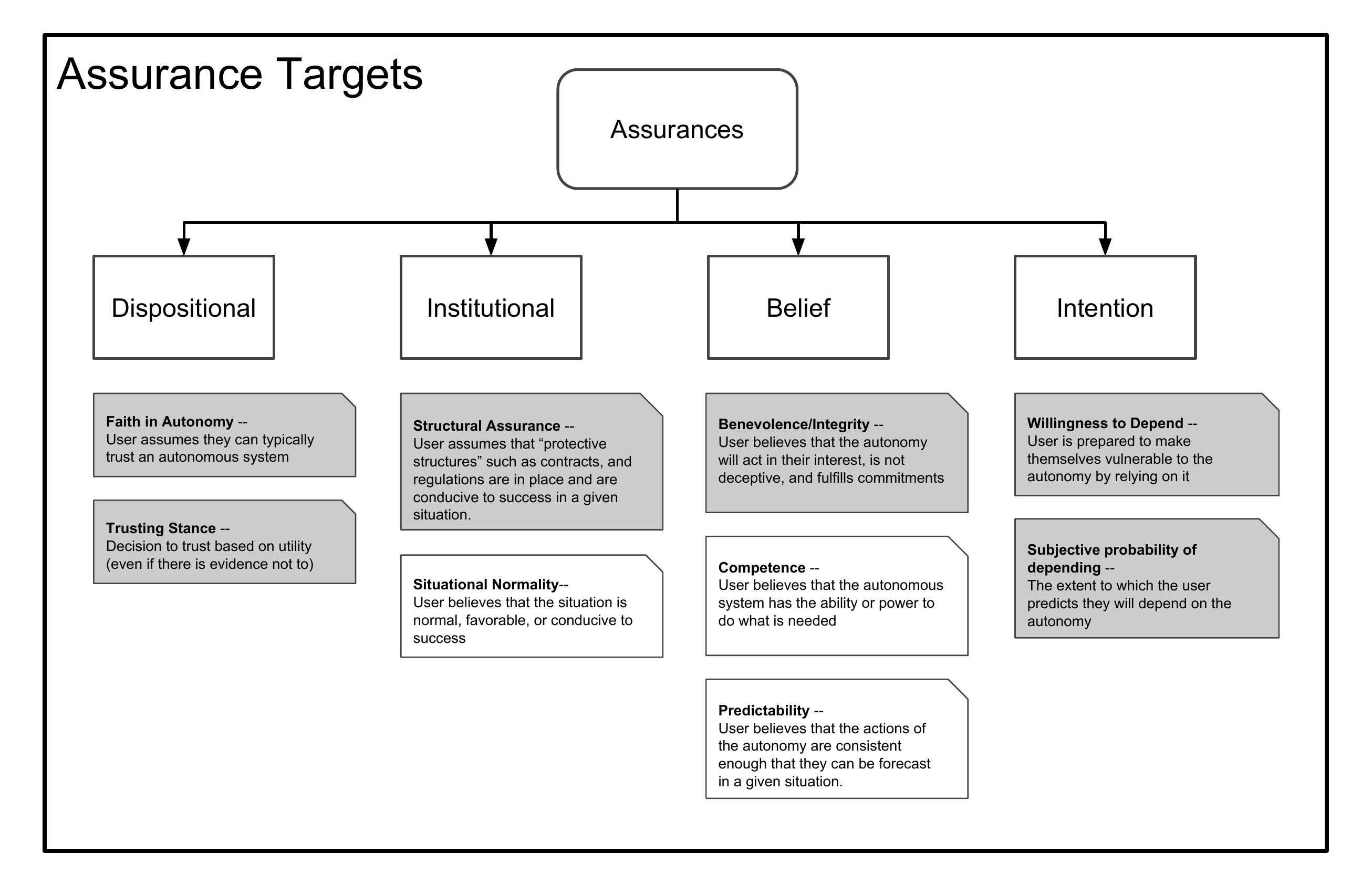}%
            \caption{Assurance targets based on the component definitions of the main categories of trust: `Disposition to Trust', `Institution-Based Trust', `Trusting Beliefs', and `Trusting Intentions'}
            \label{fig:Assurance_classes}
        \end{sidewaysfigure}

        Each of these main categories of trust has components defined in Figure~\ref{fig:Assurance_classes}. These components were defined through the compilation of many research studies across research disciplines, and because of this represent the most accurate notion of the components of trust available. These categories comprise the principal drivers of TRBs, and as such are the targets at which assurances must be directed.

\subsection{Trust-Related Behaviors} \label{sec:trbs}
Something that is well accepted among researchers of all disciplines is that trust ultimately leads to some kind of behavior or action; this idea was highlighted by \citet{Lewis1985-pr}.  \citet{McKnight2001-fa} call these `trust-related behaviors` (TRBs), which is the term that will be used in this survey. In the case of a human-AIA relationship that the author is concerned with, TRBs could include the kinds of tasks the human user assigns to the UGV such as accepting and following through on its plan, or directing that a new plan be made.

\subsubsection{Calibration of Trust-Related Behaviors}
    Trust is not a quantity that can be objectively measured. Rather, its relative magnitude must be observed through changes in TRBs, or qualitative self-reports reported in surveys \cite{Muir1996-gt}. It comes as no surprise that TRBs are the more objective measure due to the fact that people are not always consistent in their ratings, and may sincerely feel different levels of trust while performing similar TRBs. \citet{Parasuraman1997-co} were interested in understanding the use of automation by humans, and defined terms to describe that use. Here it is proposed that, by extension, those terms also apply to the behaviors of humans towards more advanced AIAs. Within this scope the definitions are as follows:
    
    \begin{description}
        \item [Misuse:] The over-reliance on an AIA (which could manifest itself in expecting too much accuracy from and AIA)
        \item [Disuse:] The under-utilization of and AIA (which could be manifest in a user turning off the AIA, or failing to use all of its capabilities)
        \item [Abuse:] Inappropriate application of automation (where \emph{application} in this case means the choice to deploy an AIA in a certain context, such as the choice to use a quad-copter underwater).
    \end{description}

    Recall the diagram in Figure~\ref{fig:SimpleTrust_one_way}; the AIA has influence on the user's TRBs by way of assurances. We propose that the AIA's goal should be that the user should not misuse, disuse, or abuse it. Consider a space of all TRBs towards an AIA, this space would include misuse, disuse, abuse, and appropriate use (all TRBs not in misuse, disuse, or abuse). In order to ensure that humans use AIAs appropriately, it is critical that the user TRBs be calibrated to elicit behaviors that are within the set of appropriate behaviors. This can only be done by influencing the user trust. This is a point that, to some extent, has been informally mentioned in \citet{Muir1994-ow,Muir1987-mk,Lillard2016-yg,Lee2004-pv,Hutchins2015-if}.

    A critical oversight of other researchers who mention `calibration' (or other synonymous concepts) is that they suggest calibrating \emph{trust} as opposed to TRBs. \citet{Dzindolet2003-ts} studied the effect of performance feedback on user's self-reported trust, and found that it increased; however the appropriate TRBs toward the system did not reflect the level of self-reported trust. This shows the danger of calibrating `trust', as opposed to calibrating the TRBs.

    Calibrating TRBs focuses on concrete and measurable behaviors that are universally applicable. In contrast, calibrating trust involves influencing a quantity that is directly immeasurable, and that, when measured indirectly, is subject to the biases and uncertainties of humans, along with inherent differences between different users. Viewing the task from this point of view, the findings of \citeauthor{Dzindolet2003-ts} are not surprising.

    It is desirable for AIAs to be designed in order to encourage appropriate TRBs, as opposed to the alternative of purposefully misleading users misuse or abuse. There is a valid argument that many of today's AIAs that ignore (or whose designers ignored) TRBs and assurances can be `unwittingly malicious' in that they do not actively attempt to guide user's TRBs to lay within the space of appropriate TRBs.

\section{Methodology} \label{sec:methodology}
    This survey examines research of those who are formally and informally investigating human-AIA trust. In particular, attention is devoted to ideas that are applicable to the trust relationship between a single human user and a single AIA. While theoretically a two-way trust model could be considered (i.e. in which the AIA also has trust in the user), attention is restricted here to a one-way trust relationship that considers only how user trust (and TRBs) evolves in response to assurances from the AIA. 

    It should be noted that it is almost impossible to perform a fully comprehensive survey of all AIA assurances, due to the broad spectrum of possible assurances, and AIAs in general. One could rightly argue that control engineers treat metrics like gain and phase margins as assurances for automatic feedback control systems, in much the same way that  machine learning practitioners treat training and test accuracy as assurances for learning algorithms -- and hence concepts related to robustness, stability, etc. for feedback control systems ought also be included in this survey. While assurances can be used in both the most simple `automatic' systems (like a thermostat), this survey will focus on assurances in more advanced AIAs that make decisions under uncertainty; however, the admittedly narrow scope of this survey does not impede the development of fundamental insights and principles in designing assurances.

    Initially, in order to find applicable research, papers that formally addressed trust, and tried to create models of it were investigated; this with the aim of trying to understand how trust might be influenced. Secondly, we researched some historical literature regarding trust between humans and some form of non-human entity (typically technological in nature). This mainly led to fields like e-commerce, automation, and human-robot interaction. Third, we looked at work regarding `interpretable', `comprehensible', `transparent', `explainable', and other similar types of learning and modeling methods. Finally, with that literature as a background, we searched for research disciplines investigating computational methods that would be useful as assurances, but in which trust itself is not the main focus.

    This information is then used to construct an informed definition and classification of assurances based off of empirical information of methods that are currently in use, or being investigated. In doing so several ideas for future research are identified.

\section{A Survey of Assurances} \label{sec:survey}
Now that AIA, trust, TRBs, and assurances have been defined we are ready to begin the survey of assurances. There are many different ways in which this survey could be organized, we choose to present it based on the different goals of the main groups of researchers who have been working in the area.

Early in reading the related literature it became clear that there were two main groups: 1) those researchers who have formally addressed the topic of trust between humans and AIAs of some form, and 2) a much larger body of those who have informally considered trust in their work (or concepts related to trust). Here we consider formal treatment of trust to include those who acknowledge a human trust model and who gather data from human users in order to measure the effect of assurances on trust. Informal treatment of trust includes those who reference the concept and/or components of trust, but who do not gather user data to verify the effects of proposed assurances. 

Another way that the landscape of researchers might be divided is by the kinds of assurances they investigate. The first group consider what we call `implicit' assurances. Implicit assurances embody any assurances that are not deliberately designed into the AIA to influence trust or TRBs. The second group consider `explicit' assurances, which are assurances that were explicitly created by a designer with the intent of affecting a user's trust. Implicit assurances can be thought of as side-effects of the design process; for example HAL 9000 could have been designed with a circular `red-eye' looking sensor because it was cost-effective, however it is possible that users who interact with HAL might find the `red-eye' sensor to suspicious, and thus lose trust in HAL. Conversely, the same `red-eye' may have been explicitly designed and selected based on several studies that indicated that users find it easier to trust advanced AIAs with `red-eye' sensors instead of similarly shaped green sensors.

Much of the research that formally considers trust has focused on implicit assurances. This is likely due to the focus on investigating what properties of an autonomous systems can affect a user's trust. It is possible to argue that someone who finds that reliability affects a user's trust is investigating an explicit assurance, but for the purposes of this paper we try to stay true to the intent of the researcher when performing the work. More recently, as seen by a large spike in interest in `interpretable', and `explainable' AIAs in government, academic, and public circles, we have seen an emergence a group who acknowledge that the concept of trust in human-AIA relationships, and who want to design systems accordingly.

\begin{figure}[htbp]
    \centering
    \includegraphics[width=0.8\textwidth]{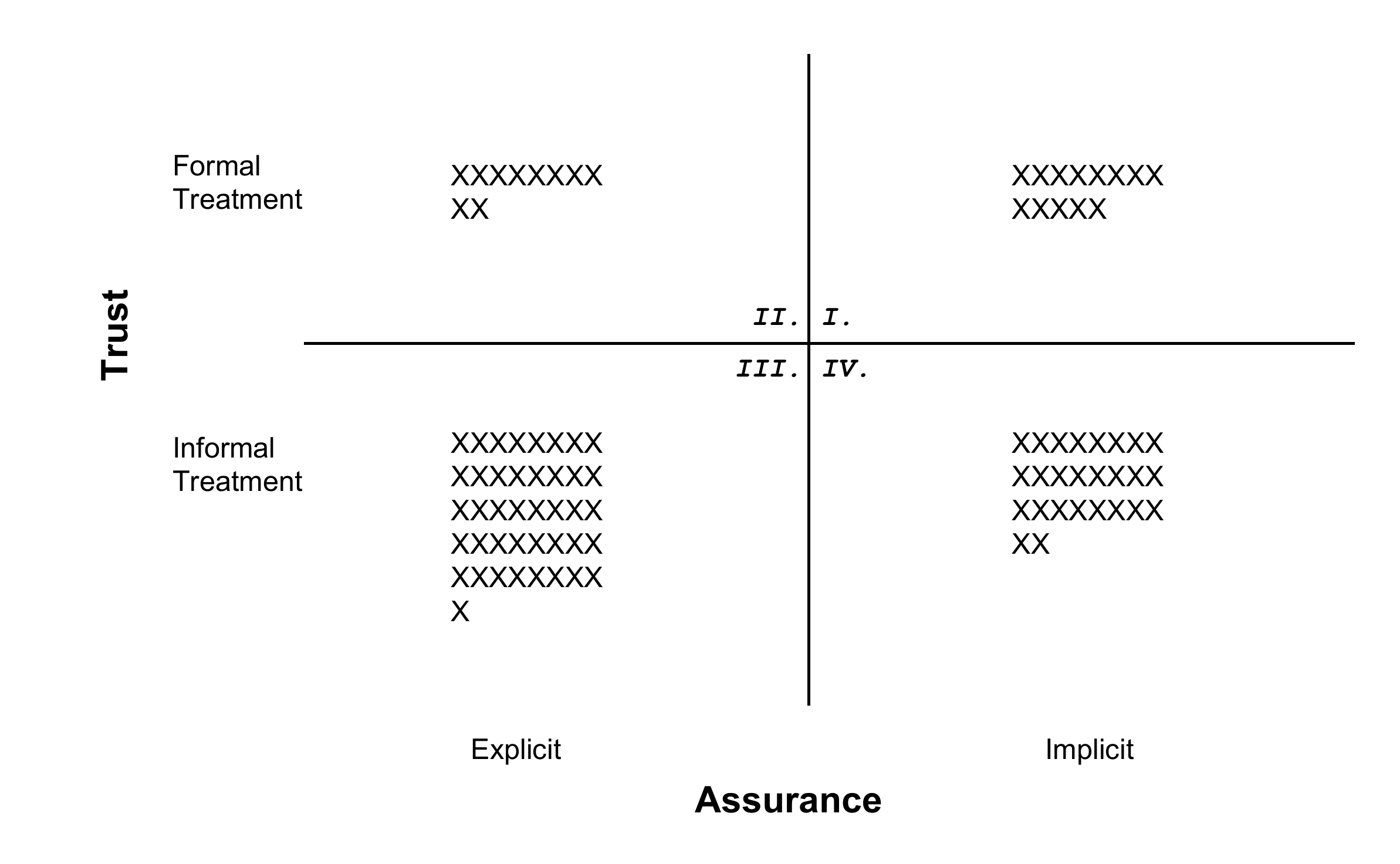}
    \caption{Figure depicting how many papers consider trust both formally and informally, as well as those who investigate explicit and implicit assurances}
    \label{fig:trust_assurance_intention}
\end{figure}

In view of these four main groups of researchers, we organize the survey by creating four quadrants shown in Figure~\ref{fig:trust_assurance_intention}. In the remainder of this section we survey each of these quadrants separately in order to gain some understanding of the lessons that each has to offer when we consider the design of assurances.

\begin{itemize}
    \item Quadrant I. (implicit assurances, formal trust treatment) -- Gather user data, consider a trust model, consider assurances that are implicit (i.e. those who care about human-AIA trust, but aren't designing assurance algorithms)
    \item Quadrant II. (explicit assurances, formal trust treatment) -- Gather user data, consider a trust model, consider assurances that are explicit (i.e. those who formally acknowledge human-AIA trust, and design assurances to affect it)
    \item Quadrant III. (explicit assurances, informal trust treatment) -- Do not gather data from users, reference trust (or its components interpretability, etc..), consider assurances that are explicit (i.e. those who know that the concept of `trust' is important, but that only use an informal notion of it when designing assurances)
    \item Quadrant IV. (implicit assurances, informal trust treatment) -- Not interested in affects on user trust, but reference (possibly only allude to) concepts that are related to trust as defined in this paper. Investigate approaches for creating AIAs with improved properties or characteristics. This work is subtly different from that in Quadrant III in the degree/intent to which trust concepts were considered. In Quadrant III trust components were clearly the main focus of the research, in this quadrant the relationship to trust is only visible to someone who knows what they are looking for (i.e. those whose work is relevant for designing assurances, but don't know it)
\end{itemize}

\subsection{Quadrant I. (Implicit Assurances, Formal Trust Treatment)}\label{sec:q1}
\citet{Muir1996-gt} performed an experiment where participants were trained to operate a simulated pasteurizer plant. During operation they were able to intervene in the fully-autonomous system if they felt is was necessary to obtain better performance. Trust was quantified by self-reported questionnaire responses, as well as by the level of reliance on the automation during the simulation. She noted that operators could learn to trust an unreliable system if it was consistent. The participants were only able to observe the reliability of the pump (i.e. the performance of the pumps over time, from which the user created a mental model of reliability).

In experiments involving thirty students and thirty-four professional pilots, \citet{Riley1996-qm} investigated how reliability and workload affected the participant's likelihood of trusting in automation. Two simulated environments were created to this end. First was for participants to use/not use an automated aid (with variable reliability) to classify characters while also performing a distraction task. Interestingly, they found that pilots (those with extensive experience working with automated systems) had a bias to use more automation, but reacted similarly to students in the face of dynamic reliability changes. In this setting, the bias to use more automation would be known as `framing effects' (where a human's trust is biased by the trust they have in previously encountered systems) in cognitive science. Findings also showed that the use of automation is highly based on individual traits.

Also considering the performance of pilots, \citet{Wickens1999-la} investigated the effect of semi-reliable data while piloting a plane. They also investigated semi-reliable performance of a system that highlighted important data for the pilot to see. The pilots were aware that the measurements/highlighting system might be inaccurate before the experiment. The reliability of the systems did have an effect on the outcome of the experiment, but interestingly did not make a measurable effect on the pilot's self-reported trust. This underscores the point that TRBs ought to be the focus of assurances as opposed to trust itself, since trust is a subjective measure that may or may not actually change a person's TRBs.

McKnight and collaborators have spent significant time investigating trust between humans and technology. His initial research was focused on e-commerce settings but later moved to trust between humans and technology. In \cite{Mcknight2011-gv} they gather self-reported trust through a questionnaire. Their experiment was interested in identifying the dimensions of trust effected by learning to use Excel for use in a business class. The results were based solely on the intrinsic properties of excel and how each individual perceived them.

In \cite{Lankton2008-ct} and later in \cite{Tripp2011-rx} they investigate the difference in trust between humans and trust between a human and technology. They found that as the technology becomes more `human-like' the self-reported trust has more similarities to trust between humans. This study was performed using Microsoft Access, a recommendation assistant, and Facebook. Respondents were asked to rate how each software `kept its commitments' and how `human-like' it was. Again, these impressions were based solely on the intrinsic properties of each of the three AIAs used in the experiment.

\citet{Freedy2007-sg} studied how `mixed-initiative teams' (MITs, their term for human-robot teams) might have their performance measured. The premise of the work is that MITs can only be successful if ``humans know how to appropriately trust and hence appropriately rely on the automation''. They explore this idea by using a tactical game where human participants supervised an unmanned ground vehicle (UGV) as part of a reconnaissance platoon. UGVs had three levels of capability (low, medium, high), and had autonomous targeting and firing capability which the operator needed to monitor in case the UGV could not perform as desired. Operators were trained to recognize signs of failure, and to only intervene if they thought the mission completion time would suffer. Trust was formally acknowledged in this survey and was quantified by using Relative Expected Loss (REL), which is the mean expected loss of robot control over $n$ trial runs. Operators were found to be more likely to use a `medium' ability UGV if they had first encountered a `high' ability UGV, as opposed to encountering a `low' ability UGV first, which is another manifestation of framing effects like \cite{Riley1996-qm}. Similar to \cite{Muir1996-gt} the operators learned to trust a UGV with low competence as long as it behaved consistently.

In a similar vein \citet{Desai2012-rc} investigated the effects of robot reliability on the trust of human operators. In this case a human participant needed to work with an autonomous robot to search for victims in a building, while avoiding obstacles. The operator had the ability to switch the robot from manual (teleoperated) mode, to semi-autonomous, or autonomous mode depending on how they thought they could trust the system to perform. During this experiment the reliability of the robot was changed in order to observe the effects on the operator's reliance to the robot. Trust was measured by the amount of time the robot spent in different levels of autonomy (i.e. manual vs. autonomous), and it was found that trust changed based on the levels of reliability of the robot.

\citet{Salem2015-md} investigated the effects of error, task type, and personality on cooperation and trust between a human and robot. In this case the robot was a domestic robot that performed tasks around the house. A human guest was welcomed to the home and observed the robot operating on different tasks. After this observation (in which the robot implicitly showed competence by its mannerisms and successes/failures) the human participant was asked to cooperate with the robot on certain tasks. Interestingly, it was found that self-reported trust was affected by faulty operation of the robot, but that it didn't seem to have an effect on whether the human would cooperate on other tasks. This seems to suggest that the effect of institutional trust (i.e. this robot may not be competent, but whoever designed it must have known what they were doing) allowed users to continue to cooperate with a faulty system, even if they have low levels of trust in it.

\citet{Wu2016-ei} use game theory to investigate whether a person's decisions are affected by whether they believe they are playing a game against a human or an AI. This idea was studied in the context of a coin entrustment game, in which trust is measured by the number of coins a participant is willing to lose by putting them at risk of the other player. On the surface, their work is meant to be a study of the implicit differences in trustworthiness between humans and robots; however in their experiment the `human' was actually an AI with some programmed idiosyncrasies to lead the human player to believe the AI was a human. This was done by adding a random wait time, as opposed to an instantaneous move that the AI would make. There were also prompts at the beginning of the `human' version of the experiment that suggested that the participant was waiting for another human player to join the game. The experiment found that humans trust an AI more than they trust a `human'. The authors suggest that this may be due to the perception that an AI does not have feelings and is operating in a more predictable way. Given that the `human' was an algorithm as well, this experiment shows that consistency (i.e. no variable wait times) was a factor that affected the trust of the participant.

\citet{Bainbridge2011-pl} investigated the difference in trust between a human and a robot, in cases where the robot was physically present and where the robot was only displayed on a screen (i.e. not physically present). In this experiment, the only method of communication from the robot was through physical gestures. Trust was measured by the willingness of the human participants to cooperate with the robot. Among other interesting findings regarding how the participants interacted with the robot, it was found that participants were more likely to cooperate with the robot when it was physically present. \citeauthor{Bainbridge2011-pl} suggest that this is due to increased trust, in this case cooperation is a TRB.

With the aim of understanding how individuals currently trust autonomous vehicles, \citet{Munjal_Desai2009-en} performed a survey of around 175 participants. The participants were asked to rate their level of comfort with six different situations. These situations ranged from parking your own car, having an autonomous vehicle with manual override park the car, and having a fully autonomous vehicle that could not be overridden park the car. There were also questions related to user comfort with autonomy in situations where they still retained control, like how comfortable users would be with having autonomous vehicles park near their car. The survey found that the participants were most comfortable with parking their own car, and least comfortable with having a fully autonomous vehicle (with no manual override) park their car. These findings are supposedly related to institutional trust, as those surveyed did not necessarily have any experience with autonomous vehicles.

\subsubsection{Summary}
We see that there have been several experiments that have formally shown the effect of implicit assurances (assurances that were not purposefully designed to affect trust) on a user's trust towards an AIA. Generally, implicit assurances can affect any of the three trust dimensions highlighted in Figure~\ref{fig:Assurance_classes}. To use a practical example recall that reliability (or rather the perception of reliability built over contiguous observations of performance) was frequently investigated as an assurance in this section. A reliable AIA can seem more competent, and predictable to a human user. Currently, it isn't very clear how different assurances affect the trust dimensions. \Citet{Muir1996-gt} attempted to identify the effects of reliability on the user's perception of the AIAs competence and predictability, but only six participants were used, so the results are questionable. Quantifying the effects of assurances on different dimensions of a user's trust is still an open research question.

Throughout this work we see evidence for the idea that a user will gather assurances, whether these are implicit or explicit, in order to execute TRBs. To restate the point, in the absence of explicitly provided assurances (such as investigated by every research paper in this quadrant) a user will still use other perceived properties and behaviors and gather assurances in order to inform their trust-related behaviors.

Even if an AIA has the ability to calculate an assurance, it must still have a way by which to express that assurance to a human user. The human user then perceives the assurance, perhaps through interaction with the system, or only through more passive observation of the system. These kinds of perceptions can be based on displayed information, or on how the AIA `behaves' (as in \cite{Salem2015-md}). Once the user perceives some kind of assurances (perhaps not purposefully communicated), those assurances are integrated into the trust of the user towards the AIA. In most cases this group of research focuses on assurances given through visual cues.

We also see evidence that human cognitive limitations need to be taken into account when designing assurances. This was directly observed in \cite{Freedy2007-sg,Riley1996-qm} where framing effects biased operator's behaviors towards the AIAs. This also suggests that other cognitive biases such as `recency effects' (being biased based on recent experience), and others will also apply as well. A couple of examples include well known cognitive biases such as `focusing effects' (being biased based on a single aspect of an event), or the `normalcy bias' (refusal to consider a situation which has never occurred before). Finally, humans naturally attempt to construct statistical models (albeit not especially accurate ones) of the world around them in order to predict and operate within it. In light of this designers must also consider the limitation (time and otherwise) for humans to build statistical models, such as reliability, when only instance by instance data is available. These ideas still remain to be verified by further, more focused, experimentation.
%

\subsection{Quadrant II. (Explicit Assurances, Formal Trust Treatment)}\label{sec:q2}
\citet{Sheridan1984-kx} were perhaps the first to attempt a formal approach to understanding the relationship between an operator/designer and a supervisory control system, such as a control system for a petroleum refinery. They considered psychological models of the human user and asked questions about how the user could understand how such systems worked. To this end they suggest that aspects of the control system need to be made transparent to the user so that the user has an accurate model of the system, but they don't offer concrete ways in which this can be accomplished. They mention that control displays should be designed according to the \emph{designer's} level of trust in the individual elements, but they seem to overlook the critical nature of what explicit assurances should be displayed to the \emph{operator} of the system, as opposed to what information might allow the \emph{designer} to have appropriate trust in the control system.

Muir investigated explicit assurances that automation could give to human operators in order to affect trust. She began by investigating formal models of trust between humans and then extending those concepts to trust between humans and automation. In \cite{Muir1987-mk} and \cite{Muir1994-ow} she investigated how decision-aids could be designed in order to affect trust. Within the framework of a trust model she suggested that a user must be able to ``perceive a decision aide's trustworthiness'', which could be accomplished by providing data regarding the system's competence on a certain task. She also suggests that summary information regarding the system's reliability over time would help. Finally she suggests improving the ``observability'' of the machine behaviors so that the user can understand it more fully.

She goes on to suggest that the user must have an accurate criteria for evaluating the system's trustworthiness. This would involve understanding the domain of competence, the history of competence, and criteria for acceptable performance of the system. She also suggests that a user must be able to allocate tasks to the system in order to feel equal in the trust relationship. The idea is that a relationship in which only one entity makes decisions is not amenable to calibrated trust. Finally she suggests that it is important to identify and `calibrate' areas where a user may trust the system inappropriately. One key shortcoming of her work is that she suggests types of explicit assurances, but does not suggest concrete approaches to implement them, or test them by experimentation.

\citet{Kaniarasu2013-ho} examine whether or not misuse and disuse (over and under trust, in their terminology) can be detected and then calibrated (aligned) to be within the AIA's competence. They use data from an experiment of a robot with a confidence indicator in a user interface. In this case the indicator was `high', `neutral', or `low' depending on the experimental mode the robot was operating in (which was unknown to the user except by the confidence indicator). A user was asked to provide trust feedback every twenty seconds while operating a robot along a specified path without hitting obstacles and responding to secondary tasks. The user trust was quantified by measuring how frequently they switched between partially autonomous and fully autonomous modes. The experiment found that the indicators of confidence did in fact reduce misuse and disuse. However, we note that at this point in time most robots are not equipped to `know' how reliable they are. This highlights a common shortcoming in the design of current robots: that they cannot quantify their own reliability.

Although they don't perform any formal experiments, \citet{Chen2014-dk} lay out a framework for agent transparency based on formal trust models. This is in the setting of situational awareness (SA, \cite{Endsley1995-ie}) of an autonomous squad member in a military operation. The aim is to make the mission SA and agent more transparent, so that the user will be able to trust the agent and use its assistance. They propose explicit feedback that can support the three levels of an operator's SA. They call their model the SA-based Agent Transparency (SAT) model. The first level -- Basic Information -- includes understanding the purpose, process, and performance of the system. The second level -- Rationale -- includes being able to understand the reasoning process of the system. The third level -- Outcomes -- involves being aware of future projections and potential limitations of the system. They suggest that two main components are display of information, and the display of uncertainty, but note that there are many considerations to take into account when trying to communicate information to human users. For example, numerical probabilities can be confusing and may need to be replaced by confidence limits or ranges. This work indicates the importance of different levels of assurances; in some situations only a certain subset of information will be needed. A key limitation is that, generally, not all of the elements of the SAT framework are available to AIAs at this time, this can be due to design limitations as well as theoretical limitations (i.e. no method might exist to quantify the future performance of a model in an environment in which it has never been deployed).

\citet{Dragan2013-wd} investigated how a robot could move in a `legible' way, or in other words, make movements that in themselves convey the intended destination. This kind of problem is important for situations in which a robot and person are collaboratively working in close proximity to each other. They investigate this within the context of how quickly a human participant is able to predict the goal of a moving point, before the point actually reaches that goal. They found that legible motion does in fact improve the user's ability to understand and predict where the robot is trying to move. It is difficult to classify this work because it does not directly address or consider human trust, but it is clearly an explicit assurance making the AIA motion capability more predictable. Furthermore, they run human experiments in order to validate that the calculated motions are in fact more interpretable to users. Their work focuses mainly on the premise that some deviation from the most cost optimal trajectory makes motion more legible; this idea can be extended in certain situations where humans rely on redundant or non-optimal behavior to predict outcomes, although it is contrast to \citet{Wu2016-ei}.

In a similar vein, \citet{Chadalavada2015-wx} investigated ways to make interaction between humans and robots more natural (i.e. being more predictable and using common human methods for communicating especially non-verbal communication) in settings where they occupy the same work space. Their approach was to have the robot project its intended movements onto the ground in order to indicate where it would be moving. They performed experiments in which participants were asked to answer questionnaires regarding how predictable, reliable, and transparent the robot was when it was projecting its intentions, and when it wasn't. There was a significant improvement in all measures when the robot was projecting its intended movements. This is strong evidence for an assurance aimed at predictability. Similarly, in \citet{Szafir2014-ok,Szafir2015-iy} they investigated using a quad-copter's motion patterns, as well as signaling with light, to help users more easily interpret the intended movements and actions.

Turning to the question of how explanations of robot reasoning can effect human-robot trust, \citet{Wang2016-id} performed an experiment in which a human and robot investigate a town; the robot has sensors that detect danger, and will recommend that the human wear protective gear if it senses danger. However, the human can choose to not accept the recommendation based on their trust in the robot, and the need to avoid the delay associated with using the protective gear. The robot is able to pose natural-language explanations for its recommendations, as well as report its ability. This was done by creating explanations generated by translating the components of the robot's planning model (the robot uses a partially observable Markov decision process (POMDP) planner) to natural language. In their experiment the robot was able to translate every component of it's POMDP which are: states, actions, conditional transitions, rewards, observations, and conditional observation probabilities. This allowed it to make statements like: ``I think the doctor's office is safe'', or ``My sensors have detected traces of dangerous chemicals'', or ``My image processing will fail to detect armed gunmen 30\% of the time''. The translation process required pre-made templates to map natural language, for example writing that observation 1 means dangerous chemicals.

During the experiment they used two levels of robot capability: `high', and `low'; and three levels of explanations: `confidence level', `observation', and `none'. They found that when the robot's ability was low, the explanations helped build trust. Generally, confidence level, and observation explanations had a similar effect on trust and both were an improvement over no explanations. Whereas, when the capability was high, the explanations didn't have a significant effect. This suggests that in some cases some kinds of assurances are overridden by other `stronger' ones, specifically the explanations (explicitly designed for affecting trust) were rendered useless by the high reliability (which can be thought of as an implicit assurance in this case) of the AIA.

Also examining POMDP-based robot planning and actions, \citet{Aitken2016-fb} and \citet{Aitken2016-cv} consider a formal model of trust, and propose a metric called `self-confidence' that is an assurance for UGV that is using a  POMDP planner (in this case, for the road network application described earlier in Section \ref{sec:mot_example}). It is made of a combination of five component assurances: 1) Model Validity, 2) Expected Outcome Assessment, 3) Solver Quality, 4) Interpretation of User Commands, and 5) Past Performance. The components considered are fairly general, and applicable to most planners, but would require new algorithms to be designed for those methods. Model validity attempts to quantify the validity of a model within the current situation. The expected outcome assessment uses the distribution over rewards to indicate how beneficial or detrimental the outcome is likely to be. Solver quality seeks to quantify how well a specific POMDP solver is likely to perform in the given problem setting (i.e. how precise the solution can be given a POMDP description). The interpretation of commands component is meant to quantify how well the objective has been interpreted (i.e. how sure am I that I was commanded to move forward). Finally past performance, is meant to add in empirical experience from past missions, in order to make up for theoretical oversights.

Self-confidence is reported as a single value between $-1$ (complete lack of confidence), and $1$ (complete confidence). A self-confidence value of $0$ reflects total uncertainty. Each of the component assurances would be useful on its own, but the composite `sum' of each factor is meant to distill the information from the five different areas, so that a (possibly novice) user can quickly and easily evaluate the ability of the robot to perform in a given situation. Currently, only one of the five algorithms (Expected Outcome Assessment) has been developed, but there is continuing work on the other metrics. As of the writing of this document, no human experiments have been performed to validate the usefulness of the self-confidence metric. 

\subsubsection{Summary}
The research in this section focuses on \emph{what} needs to be expressed as an assurance as well as \emph{how} to express it. One obvious take-away is that no AIA can express an explicit assurance unless it is able to first calculate it, or if it has been designed with assurances (i.e. a sticker on the outside saying `trust me', although this would probably be a weak assurance). Not all situations require the same kind of assurances, as discussed by \cite{Chen2014-dk}, depending on the task at hand different assurances, such as high-level predictions of future performance or confidence in executing a motion, will be more appropriate.

As in the first quadrant, we observe the critical role that human perception has on the effect assurances. Simply, an assurance can only be effective if a user can perceive it. To this end the researchers investigate different methods for expressing assurances. This methods include visually displaying `summary data' \cite{Muir1996-gt}, displaying intended actions \cite{Chadalavada2015-wx}, making movements legible \cite{Dragan2013-wd}, and using natural language in order to aid the users in perceiving the assurance \cite{Wang2016-id}.

Using explicit assurances is, unknowingly to some, an attempt to directly influence the dimensions of trust from Figure~\ref{fig:Assurance_classes}. Understanding how assurances affect the user's ideas about the `competence', `predictability', and `situational normality' of an AIA in a certain situation is an important consideration, but perhaps only in certain situations. Here researchers used metrics for tracking changes in the TRBs of a human user in lieu of attempting to understand the self-reported level of trust of the user, for example by measuring the frequency of a user switching control from autonomous to semi-autonomous. Further consideration highlights the point that -- unless the goal of the AIA is to affect the user's self-reported trust -- TRBs are really the most appropriate metric that can be used. This is an idea that began in Quadrant I, but that became more clear in this quadrant.

Most explicit assurances that have been created so far are static -- they do not change in time or with context. For example \cite{Wang2016-id} and \cite{Aitken2016-fb} consider giving calculated assurances to a user in the form of natural language feedback and simplified analysis of the task respectively. Generally, the state-of-the-art is to use assurances that are static. Given the complexity of human-AIA trust relationships, and the diverse situations and user's involved, the extension of assurances to be more dynamic and adaptable in time, and in different situations, is an important research direction that is under-served at this point in time.

\subsection{Quadrant III. (Explicit Assurances, Informal Trust Treatment)}\label{sec:q3}
\subsubsection{Performance Prediction} \label{sec:performance_prediction}
    \citet{Zhang2014-he} are concerned with the performance of visual classification systems; particularly with the common occurrence of vision systems to fail `silently', that is with no warning. They suggest that the two possible solutions are to: 1) analyze the output of the vision system to assess its confidence, and 2) analyze the input itself in order to assess the classifier's confidence. They pursue the second, with the goal of minimizing failures as well as trying to verify the safety of a classifier. They argue that the second method is more desirable because it can be applied to any classifier. In essence they learn a model of the training error associated with different images and then use this to predict the error on test images. They demonstrate their methods on several different classification tasks and show that it predicts failure ``surprisingly reliably''. Although, they make an assumption that similar inputs (by some measure) will give similar outputs, which can be difficult to quantify in some domains.

    \citet{Gurau2016-hs} (following from ideas in \citet{Churchill2015-ei}) consider the task of predicting the performance of a robot's navigation when using cameras for localization. Their aim was to have the robot operate autonomously only when it was confident that it could do so, otherwise it would request human assistance. To achieve this they make use of GPS localized observations and `Experience Based Navigation' (EBN) to identify which past experience matches the current state of the robot. Using this approach they are able to reduce the number of mistakes that the robot makes when navigating in a real scenario. This approach seems to be fairly sound in their suggested application, and they claim that it is agnostic of classifier, which should be beneficial for use in other AIAs.

    \citet{Kaipa2015-hy} also consider the confidence of a visual classifier, but with the application being a robot that is picking an object out of a bin. To do this they utilize the `Iterative Closest Point' (ICP) method to match which points in a point cloud are likely associated with the object of interest in the bin. With that information the robot can then asses how confidently it can perform the task and if its confidence is below some threshold it will request assistance from a human. This method is quite similar to `Chi-square innovation hypothesis testing' used for adaptive Kalman filtering \cite{Bar-Shalom2001-tg}, and shares a key drawback, which is that it requires a precise model, and doesn't necessarily indicate \emph{why} the test is failing. Regardless, if applicable they are quite useful for verifying that the AIA is `OK'.

    \citet{Kuter2015-qh} recognize that plans produced by hierarchical tasking networks can be fragile, and suggest that the planner calculate its stability. In essence, how sensitive is the plan to uncertainties? They suggest `counter planning' and `plan repair' so that the autonomous system can identify likely contingencies that might interfere with an existing plan and then adapt the plan to account for those contingencies. If the system \emph{is} able to adapt to contingencies then it will have a higher `self-confidence'. This allows humans to be able to understand the system better, and trust it more appropriately.

    \citet{Hutchins2015-if} consider the fact that autonomous systems have components, and that those components have `competency boundaries'. Which refers to the fact that different sensors, planners, and actuators have different capabilities in different conditions. For example a GPS has high competency in an open-air situation, but very low competency inside of a tunnel or building. They suggest that these competency boundaries be quantified, and then displayed to a user, so the user better understands the system's capabilities and can trust it more. The main limitation here is that they propose an expert-based Likert scale for quantifying competency boundaries, which, which  lacks formal foundations, and extensions to other domains. Both this work and that of \citeauthor{Kuter2015-qh} are related to the work of \citeauthor{Aitken2016-fb}, but the later two do not explicitly relate their approaches to trust and TRBs. Although, \citeauthor{Hutchins2015-if} is the only paper of the three to consider the exact method by which the assurance should be communicated.

\subsubsection{Interpretable Models} \label{sec:model_interp}
    A well-known concept in machine learning and pattern recognition is the trade-off between accuracy and interpretability. \citet{Van_Belle2012-dt} examine this issue in the context of making clinical decision support systems for use in the medical field. To this end they propose an `interval coded scoring' system that imposes a constraint such that each variable has a constant effect on the estimated decision-making risk (e.g. the Bayes' risk for classification).  They demonstrate the method on two datasets, and show that their method can be visualized using scoring tables and color bars. They point out that their method needs to be extended in order to detect interaction effects automatically, and to handle multi-class decision support systems. One reason why this approach is important is because probabilities are notoriously difficult for humans to interpret, and thus the reason to manipulate the output in order to accommodate those who interact with the system.

\begin{figure}[htbp]
    \centering
    \includegraphics[width=0.8\textwidth]{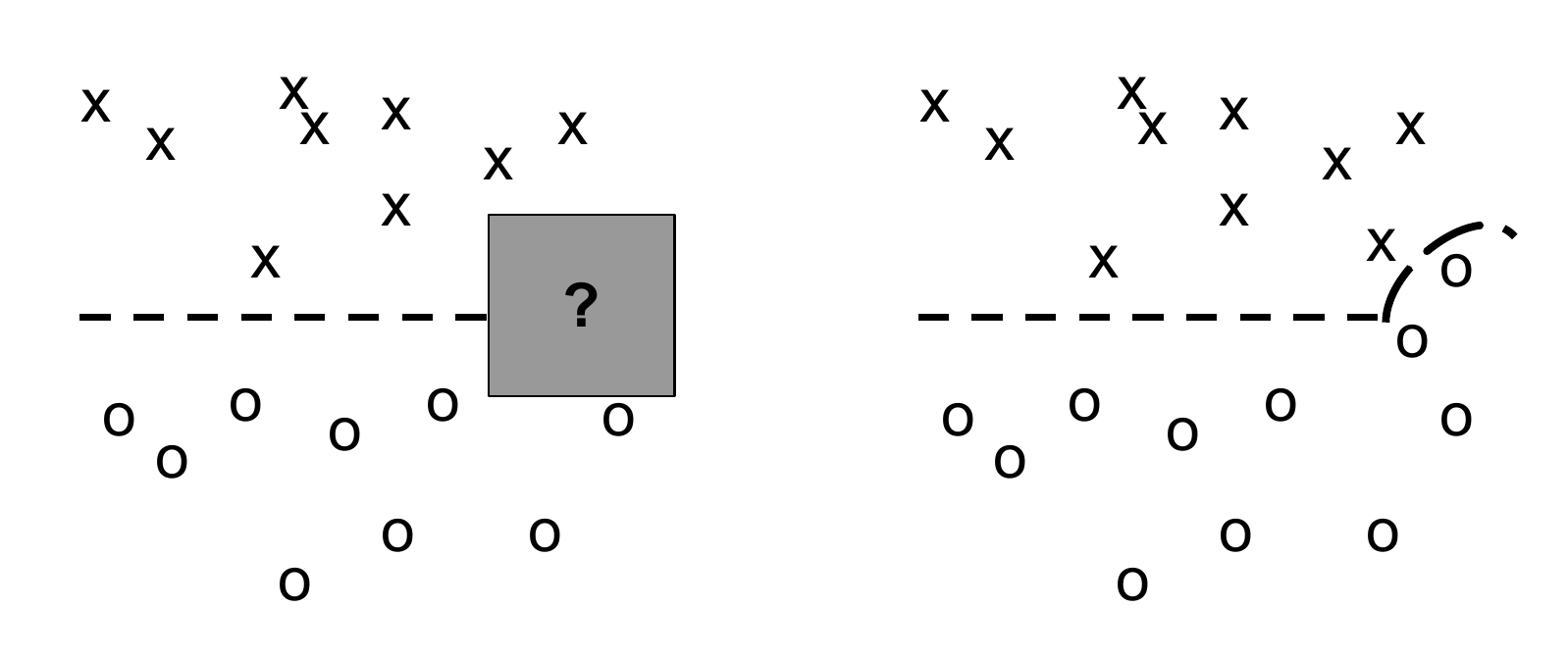}
    \caption{Example of simple global model on the left, and on the right a more complex local model that can be used for interpretability when more detail is necessary.}
    \label{fig:ruping}
    \end{figure}

    Along these lines,  \citet{Ruping2006-xj} specifically asks how classification results can be made more interpretable to those who design and use classifiers. He reviews several possible methods, and states that ratings by experts is perhaps the most accurate way, although it is at the same time subjective. To address the accuracy-interpretability trade-off he investigates the use of a simpler global model, and a more complex local model (\citet{Otte2013-oo} and \citet{Ribeiro2016-uc} implement similar ideas as well). Figure \ref{fig:ruping} illustrates this idea on a simple example, the left (global) model can be used as a large scale interpretable model, while the more complex local model can be used as necessary when more precise understanding is required. This allows simple interpretation of global properties, but more complex local models to maintain accuracy. While his focus was on classification, the methods could also be useful in regression as well. He spends time investigating how the global and local models should be learned in a principled way, and demonstrates this approach on several black-box and white-box classifiers. 

    Later, \citet{Van_Belle2013-ph} address the challenge that some of the highest performing ML models are too complicated to be interpreted. They investigate different `interpretable' and visualization methods in order to understand what opportunities they find. They suggest that there are three methods that help ascertain the level of interpretability and potential utility of models: 1) Map to domain knowledge, 2) Ensure safe operation across the full operational range of model inputs, and 3) Accurately model non-linear effects (compare to categories proposed by \citet{Lipton2016-ug}). They analyze some of the existing methods (nomograms, rule induction, graphical models, data visualization) and point out their weaknesses. They finish by pointing to more recent research in sparse and interpretable models, and suggest that it is a promising line of research. In summary, they discuss that each `interpretable' method has its benefits and weaknesses and there is no method that clearly out-performs any other.

    Similarly, \citet{Caruana2015-za} are interested in predicting the probability that a patient will be re-admitted to the hospital within thirty days. They also mention the trade-off between accuracy and interpretability, and propose a generalized additive model with pairwise interactions ($GA^2M$) model. A generalized additive model (GAM) is generally thought of as interpretable as the effects of individual variables can easily be quantified; however GAMs suffer from low accuracy. In a comparison with other methods, they show that adding pairwise interactions allows the $GA^2M$ model to be as accurate as the current less interpretable methods, while still maintaining reasonable interpretability.

    In a similar vein, \citet{Choi2016-by} modify a recursive attention neural network to remove recurrence on the hidden state vector, and instead add recurrence on the doctor visits and diagnoses. In this way the model is able to predict possible diagnoses in time, and a visualization can be that that indicates the critical visits and diagnoses that lead to that prediction. These methods are promising because it restructures an advanced learning model in a way that useful information can be extracted. This method seems very promising, however, their architecture is dependent on problems with similar properties, meaning that their approach would need to be re-implemented by an expert to be used in another situation.

    \citet{Abdollahi2016-vn} construct a conditional restricted Boltzmann machine (RBM) in order to create a collaborative filtering algorithm, that can suggest `explainable' items, while maintaining accuracy. The challenge is that the RBMs are very accurate collaborative filtering algorithms, but difficult to understand. They make a more explainable version by adding an additional `explainability layer' which indicates the explainability of an item to a certain user. This method shows how adding an additional element in a model can help add explainability, and this seems to be somewhat of a theme in this section.

    \citet{Ridgeway1998-lv} recognize that boosting methods, or classifiers with voting methods, are very accurate but not interpretable. They propose boosting by `weight of evidence' (WoE), where WoE refers to having a weight that indicates whether the observation is positively or negatively correlated to the class. Each weight can then be used to gain some understanding about how the observation affects the classification. They demonstrate its utility using multiple experiments that it has performance on par with AdaBoost, and with a Naive Bayes classifier.  Again, their approach gets at the ability of adding in interpretability components into a model can yield a more interpretable model, while losing very little accuracy.

    \citet{Huysmans2011-th} investigate decision trees, decision tables, propositional if-then rules, and oblique rule sets in order to understand which is the most interpretable and perform an experiment to identify which method works most effectively. The experiment involved interpreting a model and answering questions about what the correct classification would be. They made several interesting observations. First, they confirmed that a model with larger representation leads to lower answer accuracy responses. They found that overall decision trees and decision tables were the most interpretable, but that different tasks made the tree or table more desirable (the layout of the data in a table can be superior for certain `look-up' tasks). 

    One drawback with decision trees is that the rules can get very complicated in a large tree. \citet{Park2016-ld} investigate how to make rules in decision trees more `intuitive'. They propose a method that learns chains of decisions that together increase the ratio of positive class. They present a method that is called $\alpha$-Carving decision Chain (ACDC), and say that it is a greedy alternative to a general framework known as RuleLists (\citet{Wang2015-ww}). Similarly, \citet{Jovanovic2016-gw} use `Tree-Lasso' logistic regression with domain knowledge. Specifically they use medical diagnostic codes to group similar conditions and then use `Tree-Lasso' regression that uses that information to make a more sparse model. \citet{Zycinski2012-jj} also use domain knowledge to structure the data matrix before feature selection and classification. While the work of \citeauthor{Huysmans2011-th} gives some indication about how to choose classifiers for interpretability, \citeauthor{Park2016-ld} points out that to gain \emph{real} interpretability in complex tasks expert knowledge is still needed to make sense of complicated features.

    \citet{Faghmous2014-og} argue that `theory guided data science' is necessary in science applications; for instance, when studying environmental effects, black-box models are of little use. Instead, in order to gain insight they need a theoretical framework, to help highlight causation. Similarly, \citet{Morrison2016-fz} address the situation where an analytical model is available but imperfect. They use a chemical kinetics application where the theoretical reaction equations are well known, and then add a `stochastic operator' over the top of the known model to account for uncertainties. One drawback for these methods is that they require a lot of domain knowledge. However, in the areas of science and engineering where physical models are well understood this type of approach could be useful to indicate the performance of the system with regards to the theoretical model.

\subsubsection{Visualization and Dimensionality Reduction} \label{sec:viz_dr}
    Intuitively, it might make sense to simply show users the inputs, raw data, and/or intermediate processing steps that led to a particular AIA behavior. In many real applications, however, there are too many individual variables for a human to attend to. In such situations, dimensionality reduction (DR) and visualization are tools that can be used to help make a model or data easier to understand. \citet{Venna2007-yj} discusses dimensionality reduction as a tool for data visualization for ML, and reviews many linear and non-linear projection methods. \citet{Vellido2012-nm} also discusses the importance of DR for making ML models interpretable.

    In an interesting application to time series data, \citet{Kadous1999-rx} asks how comprehensible descriptions of multivariate time series can be learned, with the end goal of interpreting Australian sign language. He focuses on reducing the feature space fed into a learning algorithm (i.e. the data was gathered using an instrumented glove), and does so using `parameterized event primitives' (PEPs), which are commonly occurring events or patterns that are expected to occur. He shows that his method reduced test error while also having more comprehensible features. It seems likely that parameterized primitives might be automatically learned. While an interesting application, this would require expert knowledge to extend to other domains. Although it does seem likely that PEPs could be automatically identified in some way.

\subsubsection{Explanation} \label{sec:explanation}
    In the context of POMDP planning, \citet{Lomas2012-ie} recognize that in order for humans to appropriately trust robots they must be able to predict their behavior, and the robot must be able to communicate in order for that to happen. To this end they present the Explaining Robot Actions (ERA) system that interfaces with a model that represents semantic and physically based information, along with other factors. In essence the ERA is a translator between the planner/model and the human. This approach depends on a layered world model that includes semantic information as well as the physical model, but is promising in the respect that the ERA is a separate system that queries the robot. 

    With regards to expert systems \citet{Swartout1983-ko} examines explanation methods. He noted that ``trust in a system is developed not only by the quality of its results, but also by clear description of how they were derived''. Often the data (or knowledge) used to train an expert system is not kept in the system for later use. He proposed a method called `XPLAIN' that not only describes \emph{what} it did, but \emph{why} it did it. It does this by using description facts of the domain and prescriptive principles simultaneously; in essence it learns how to make decisions and how to explain them at the same time. This approach is meant for a  structured problem defined by a domain model and the domain principles, in situations where problems are based mostly in data, this approach would not be applicable.

    \citet{Rouse1986-dz} asked how computer-based decision support systems should be designed in order to help humans cope with their complexity. He suggests that methods need to be designed so as to provide different kinds of information which include: patterns vs. elements, and current vs. projected outcomes or states. This work is important in pointing out that assurances also depend on what the role of the human is and what information they need.

    This was also investigated to some extent by \citet{Wallace2001-fm} in their work concerning explaining outcomes of constraint solvers. They discuss how to distinguish between `good' and `bad' explanations, or those explanations that facilitate or detract from the user's ability to understand how the explanation actually applies to the solution being explained. They critically ask the question of how explanations should be presented to users (something that \citet{Kuhn1997-qc} explores more formally with regard to framing of uncertain information in decision making). An important consideration is that constraint solvers don't take uncertainty into account, but instead are deterministic. This is not the case in many of the more adaptable and capable AIAs.

    \citet{Lacave2002-cu} revisit some of these ideas from the perspective of explaining Bayesian networks. They are concerned with \emph{how} and \emph{why} a system reaches a conclusion. They present three properties of explanation: 1) content (what to explain), 2) communication (how to explain), and 3) adaptation (how to adapt based on who the user is). It is not possible to cover all of the ideas that they present in their paper, but they are key to the idea of designing assurances. Some key points are that they highlight the differences between explaining evidence, the model, or the reasoning. These are three key considerations in making assurances. They also discuss whether an explanation is meant to be a description or for comprehension, as well as whether explanations need to be on a macro or micro scale (as mentioned by \citeauthor{Ruping2006-xj}). They also consider whether explanations should occur by two-way interaction between system and user, by natural language interaction, or by probabilities. Finally, when considering adaptation, they hit on another key point of assurances, which is that in general application not all users will require (or desire) the same kinds of assurances. This paper points out many challenges and considerations in designing assurances, and illustrates that, as with the `No free lunch' theorem, there is no single `best' assurance that will address every possible situation.

\subsubsection{Model Checking} \label{sec:model_checking}
    While assurances behind reasoning processes can be useful in many situations, they are not trustworthy in and of themselves if the models or assumptions they are based on are flawed to begin with. Thus, there is also great interest in providing assurances that the models and assumptions underlying said AIA reasoning processes are, in fact, sound. \citet{Laskey1991-mf} -- with the intention of helping users of `probability based decision aids' by communicating the validity of the model -- notes that it is infeasible to perform a decision theoretic calculation to decide if revision of the model is necessary. She then presents a class of theoretically justified model revision indicators which are based on the idea of constructing a computationally simple alternate model and then to initiate model revision if the likelihood ratio of alternate model becomes too large. 

    \citet{Zagorecki2015-qy}, discusses the `surprise index' introduced by \citet{Habbema1976-xd}, which is the likelihood of a certain measurement given a specific model, which applies nicely to Bayes-nets that have probabilistic descriptions. The major flaw of the surprise index is that it is computationally infeasible due to the possibility (likelihood) of being a non-analytic distribution. \citeauthor{Zagorecki2015-qy} suggest an approximation by using the log-normal distribution to approximate the distribution of values in the joint probability distribution. Another challenge is knowing what a `good' value for the surprise index is since knowing the maximum likelihood of a non-analytic distribution is not an easy problem.

    \citet{Ghosh2016-dl} present a method, in the framework of a practical self-driving car problem, called Trusted Machine Learning (TML). The main approach of TML is to make ML models fit constraints (be trustable). To do this they utilize tools from formal methods to provide theoretical proof of the functionality of the system. They present `model repair' and `data repair' that they can utilize when the current model doesn't match the data, at which point the model and data can be repaired and control can be replanned in order to conform with the formal method specifications. One challenge that presents itself is how to identify the `trustable' constraints, this returns a lot of responsibility to the designer to foresee all possible failures, which is a strong assumption.

\subsubsection{Human-Involved Learning} \label{sec:human_involved}
    Another possible way to make system assure a human user is to use the human in the learning process. \citet{Freitas2006-qo} addressed this point with regards to discovering `interesting' knowledge from data, by comparing two main approaches. Given such large datasets, human users require assistance from complex systems in order to find patterns and other `interesting' insights. He mentions `user-driven' methods that involve a user pointing out `interesting' templates, or in another method general impressions in the form of IF-THEN rules. He compares these methods to other `data-driven' methods that have been used, and cites other research that suggests that data-driven approaches are not very effective in practice. This is a cautionary tale that many times engineering methods to assist humans are not as effective as we would like to believe. Although, the `user-driven' approach may not fair any better when compared over many users, as each user will likely have different preferences. \citet{Chang2017-kl} also consider a similar, scaled up, `user-driven' approach called `Revolt' that crowd-sources the labeling of images. It is able to attain high accuracy labeling, while also exploring new or ambiguous classes that can be ignored with traditional approaches.

\subsubsection{Explicit Assurances via Formal Methods} \label{sec:VV}
    Validation and Verification (V\&V) typically refers to using formal methods to guarantee the performance of a system within in some set of specifications. Not all practitioners are aware that V\&V provides ways to assure users; arguably some have this idea in their mind, but only a few V\&V researchers consider how to communicate the results of analysis via formal methods to users. A prime example is given by \citet{Raman2013-mz}, who developed a way by which a user can provide natural language specifications to a robot and a `correct-by-construction' controller will be built if the specification is valid. Otherwise, the robot will provide an explanation about which specification(s) cause failure. They study this with the goal of implementing the system on a robot assistant in a hospital. Their method involves parsing natural language input (such as: ``Go to the kitchen''), and converting that to linear temporal logic (LTL) that represents a task specification. This is then used to construct a controller if possible, otherwise the `unrealizable' specifications need to be communicated back to the user. This approach is promising in that it presents a way to communicate that a specification cannot be met, although it does not formally account for effects on user trust or TRBs in formulating explanations. The expression of assurances is also asymmetrically limited to cases where the robot cannot meet the specifications. 

\subsubsection{Summary}
    Many of the approaches above focus on informing the user how the model/logic works; there is also some focus on predicting future performance. Some AIAs do not have the capability to implement these approaches. Or, the type of approach -- such as informing the user how a deep neural network works -- simply does not have a satisfactory answer to date. As another example, methods for quantifying the amount of uncertainty in a planner have not been developed for all planning algorithms. Likewise, there may not be satisfactory ways of visualizing certain data. 

    The approach of using a more simplistic global model paired with more accurate local model is fairly prominent (\cite{Ruping2006-xj,Ribeiro2016-uc,Otte2013-oo}). This idea is similar to how human experts explain approaches to non-experts; as a non-expert desires to know more about a specific concept of a complex idea, more detailed explanation can be provided. This is promising because it is also extensible to different types of users. In particular \cite{Ribeiro2016-uc} is designed to work with any classifier, and \cite{Zhang2014-he} has the same goal as well. Of course the trade-off of accuracy for flexibility will always exist between more general approaches and specialized approaches.

    The underlying hypothesis here is that many of the methods should affect the user's perception of the `competency', and `predictability' of the AIA, or the `situational normality' of the task being performed. However, none of this has been tested by experimentation that gathers self-reported changes in trust. Similarly, there has been no formalization of how the effects of assurances on TRBs might be quantified in different applications. In essence this work is only addressing at a small subset of important considerations for designing assurances, or the subset that includes the methods by which to calculate the assurance. The experimental testing of the effects of the proposed assurances on both self-reported trust, and TRBs remains open. Indeed, the research in this section is mainly focused on what AIA engineers and designers \emph{think} needs to be explained, or what assurance they \emph{think} users should have. These ideas definitely have merit, but need to be tested to identify if they are effective, to what extent they are effective, and whether there is a more effective, or more efficient way to achieve similar results.

\subsection{Quadrant IV. (Implicit Assurances, Informal Trust Treatment)}\label{sec:q4}
\subsubsection{Safety and Learning under nonstationarity and risk aversion} \label{sec:safety}
    While a fairly high-level treatment, \citet{Amodei2016-xi} are concerned with `AI safety', which is in essence how to make sure that there is no ``unintended and harmful behavior that [emerges] from machine learning systems''. Given this definition, much of the paper discusses concepts that are critical to AIA assurances. Among the more directly applicable topics in the scope of this paper are: safe exploration (how to learn in a safe manner), and robustness to distributional shift (a difference between training data and test data). They also discuss designing objective functions that are more appropriate. To restate more concretely, there is a need to design objective functions that more accurately reflect the true objective function. A popular example (roughly summarized here) from \citet{Bostrom2014-fz} is a robot that has an objective of making paper clips, it then decides to take over the world in order to maximize its resources and ability to make more paper clips. This highlights the point that sometimes over-simplistic objective functions can result in unintended and unsafe behaviors.

    Generally, stationary data (data whose distribution does not change after training) is assumed in supervised ML. \citet{Sugiyama2013-ci} considers what to do when there is `covariate shift' (or when both training and test data change distribution, but their relation does not change between training and test phases), and `class-balance change' (where the class-prior probabilities are different in training and test phases, but the input distribution of each class does not change). They design and present tools that help diagnose and treat those conditions (this is follow on work for some of what is presented in \citet{Quinonero-Candela2009-fj} where they consider dataset shift). A key approach is to use importance sampling, which involves weighting the training loss by the ratio between the probability of the test data and that of the training data.

    \citet{Hadfield-Menell2016-ws}, in considering an AI safety problem, address the problem of verifying that a robot can be turned off, even when it might have an objective function that indicates that disabling the `off-switch' will reduce the chance of failure. This kind of scenario, or something similar, can easily occur with a sufficiently sophisticated and capable AIA, and a complex enough set of objectives that might result in unintended consequences. They propose that one objective would be for the robot to maximize value for the human, and to not assume that it knows how to perfectly measure that value. 

    \citet{Da_Veiga2012-gh} discuss a form of safety where they are concerned with nonparametric classification and regression with constraints. More specifically, they are concerned about learning Gaussian process (GP) models with inequality constraints, and present a method to do this by using conditional expectations of the truncated multivariate normal distribution (or Tallis formulas). This is not the only work that references learning with constrained GPs. It is also not the only work that considers constrained modeling, but it would take too long to review all of those papers. The main claim here is that constrained models are a way to guarantee the properties of a model within some specifications.

    \citet{Garcia2015-rs} perform a survey about safe reinforcement learning (RL). Safety in RL can be critical based on the application, such as an aerospace vehicle that can cost several thousands of dollars. They state that there are two main methods: 1) modifying the optimality criterion with a safety factor, and 2) modification of the exploration process through the incorporation of external knowledge. They also present a hierarchy of approaches and implementations. Some approaches used when modifying the optimality criterion are worst-case, risk-sensitive, and constrained criterion. Whereas, modifying the exploration process is done through using external knowledge, and as well as using risk-directed exploration. Safe RL is a particularly important area that requires assurances, as the systems are designed specifically to evolve without supervision.

    As one example, \citet{Lipton2016-dq} design a reinforcement learner that uses a deep Q-network (DQN) and a `supervised danger model'. Basically, the danger model stores the likelihood of entering a catastrophe state within a `short number of steps', this model can be learned by detecting catastrophes through experience and can be improved over time.  In this way they show that their method, they call `intrinsic fear', is able to overcome the sensitive nature of DQNs. There is a clear limitation of the danger model, in that it does not contain useful information until a catastrophe is detected. 

    \citet{Curran2016-ij}, in a more specific application, asks how a robot can learn when a task is too risky, and then avoid those situations, or ask for help. To do this, they use a hierarchical POMDP planner that explicitly represents failures as additional state-action pairs. In this way the resulting policy can be averse to risk. They say that this method can be especially useful when optimal actions are not straight-forward, and they state that it can use any reward function. It seems that this method might suffer from some of the typical problems of POMDPs, which are computational complexity in high-dimensional state spaces.

\subsubsection{Active Learning} \label{sec:active_learning}
    \citet{Paul2011-vr} is concerned with whether a robot can improve its own performance over time, with the goal of `life-long learning'. They use `perplexity' which is a method first introduced in language models and adapted to work with images. Perplexity, in their application, is a measure that indicates the uncertainty in predicting a single class. Over time, the most perplexing images are stored and used in expanding the sample set. This work is interesting for application in assurances because the ability to quantify something that is perplexing is a predecessor to being able to communicate that to a human user.

    Recently, there have been several papers that attempt to use Gaussian processes (GPs) as a method to actively learn and assign probabilistic classifications (see \citet{MacKay1992-sp,Triebel2016-kj,Triebel2013-ow,Triebel2013-ku,Grimmett2013-gj,Grimmett2016-yc,Berczi2015-rd,Dequaire2016-kh}). The applications surveyed here are all mainly related to image classification and robotics. As with perplexity-based classifiers, the key insight is that if a classifier possesses a measure of uncertainty, then that uncertainty can be used for efficient instance searching, comparison, and learning, as well as reporting a measure of confidence to users. The key property of GPs that makes them an attractive for this purpose is their ability to produce confidence/uncertainty estimates that grow more uncertain away from the training data. That is, GPs have the inherent ability to `know what they don't know', and this information can be readily assessed and conveyed to users, even in high-dimensional reasoning problems. This property of GPs has also found great use in other active learning applications, such as  Bayesian optimization (see \citet{Williams1998-kr}, \citet{Snoek2012-tt}, \citet{Brochu2010-tj}, and \citet{Israelsen2017-zb}).

\subsubsection{Representation learning and Feature Selection} \label{sec:rep_learning}
    Another promising field of research is related to learning representations of data and selecting data features. These two topics are surveyed by \citet{Bengio2013-uv} and \citet{Guyon2003-fj} respectively. From some of the discussion of interpretable models in section \ref{sec:model_interp} we find that representation is important for making interpretable models. Having appropriate representations (i.e. like the ones humans use and understand) is a large step forward in making assurances for humans.

    For instance, in their work related to interpreting molecular signatures \citet{Haury2011-zi} investigate the influence of different feature selection methods on the accuracy, stability and interpretability of molecular signatures. They compared different feature selection methods such as: filter methods, wrapper methods, and ensemble feature selection. They found that the effects of feature learning greatly influenced the results. 

    As another example, \citet{Mikolov2013-lt} studied how to represent words and phrases in a vector format. Using this representation, they are able to perform simple vector operations to understand similar words, and the relative relationships learned. For example the operation $airlines+German$ yields similar entries that include $Lufthansa$. This type of representation encodes information that can be checked and understood by humans. 
    
    How can human understandable features and representations be discovered? This is still an open question. The main question in the representation learning world is how to find the best representations, not necessarily the representation and features that are most human. This is not surprising because human representations and features are not necessarily optimal, and AIAs are being designed to be optimal using other objective functions (arguably more appropriate functions, if humans don't need to understand what is going on).

\subsubsection{Summary}
    The literature surveyed in this section is \emph{not} exhaustive, nor could it reasonably ever claim to be. One thing that should be clear is that in every field where designers want to ensure reliable and correct application of an AIA, there will be assurances that are created. The disciplines selected in this paper are a subset that are aligned with the author's interests in unmanned vehicles.

    The work in this section is easily distinguished from that in Quadrant I because it does not discuss trust in any way. However, it is only subtly different from that in Quadrant III. The research in Quadrant III is explicitly focused on things like interpretability and explanation by direct statement of the authors. Conversely, the research found in this quadrant is only related to trust by those who are familiar with the underlying concepts in this paper. This research is created with the intent of making the AIAs intrinsically more safe, aware of reasoning processes, having better representations in some way, and others. This group unintentionally, creates foundations for trustworthy AIAs. Here are found the researchers who created AIAs with properties, like reliability, that can then be investigated by those who formally acknowledge human-AIA trust in Quadrant I. These are the methods can be turned into explicit assurances by designers who intentionally do so.

\section{A Synthesized and Refined View of Assurances} \label{sec:synthesis}
\begin{figure}[htbp]
        \centering
        \includegraphics[width=0.9\textwidth]{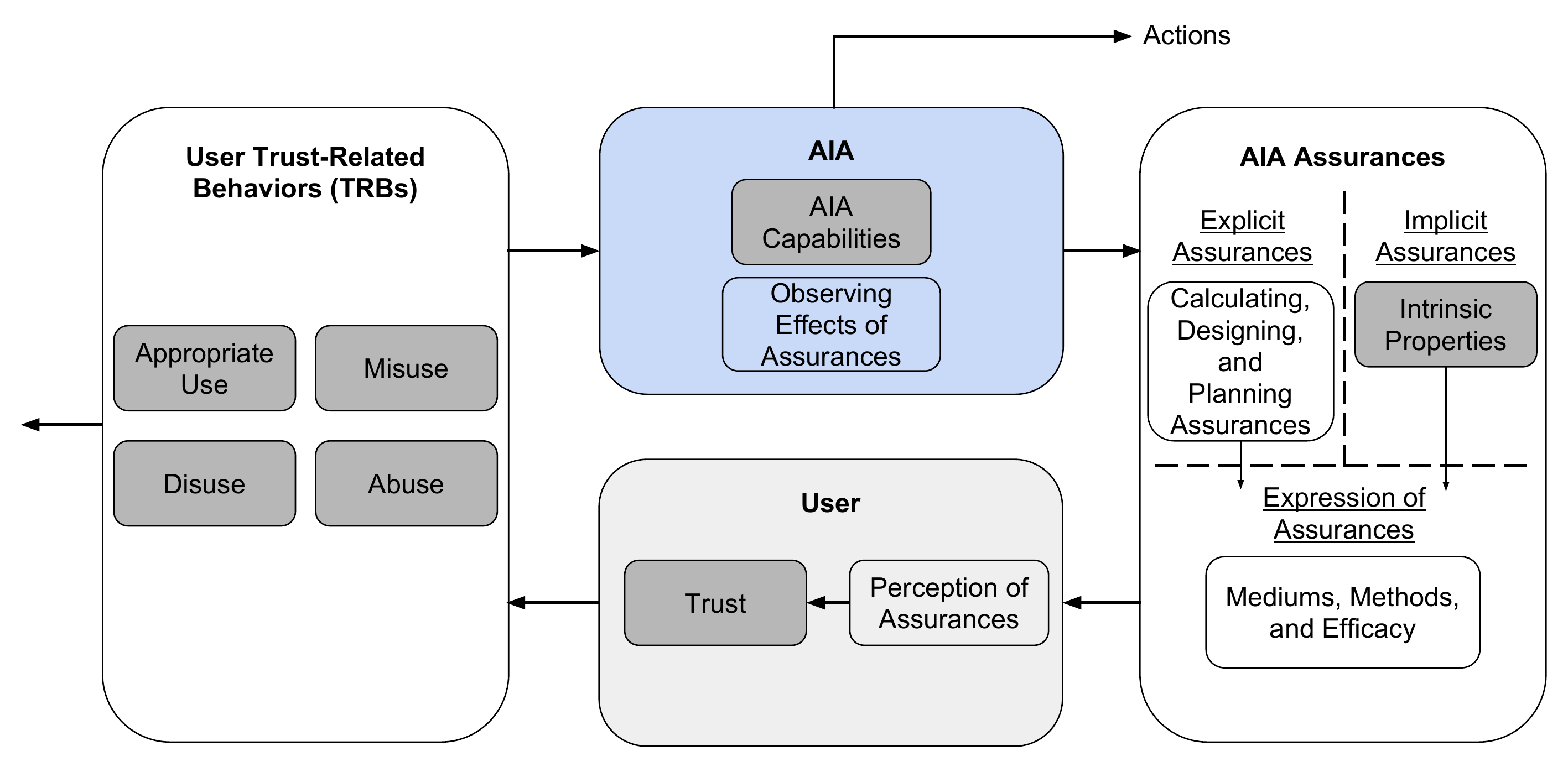}
        \caption{Detailed extension of Figure~\ref{fig:SimpleTrust_one_way}. The AIA, User , and User TRBs blocks are defined as discussed in Section~\ref{sec:background} (with the exception of the `Perception' blocks added to the AIA and User boxes). The AIA Assurances box has been filled using insights from the surveyed material. The boxes that are greyed out will not be discussed in this section.}
        \label{fig:refined_assurances}
    \end{figure}

    From the review of Quadrants I. through IV. of the formal/informal, explicit/implicit plane, we are able to find some insights with respect to assurances and can discuss them in a more comprehensive way. Using insights from the survey a refined version of Figure~\ref{fig:SimpleTrust_one_way} can be constructed. Figure~\ref{fig:refined_assurances} incorporates all details from Section~\ref{sec:background} as well as adding some insights from the survey that give direction about the design of assurances in human-AIA trust relationships. 
   
   Below we synthesize and discuss the design of explicit assurances in this more detailed framework -- some of these insights might also apply to implicit assurances, but implicit assurances will not be directly discussed in this paper.

\subsection{Calculating, Designing, and Planning Explicit Assurances}
    Recall that an assurance is defined as \emph{any} behavior or property of an AIA that affects a user's trust, therefore an explicit assurance is any assurance that was consciously implemented/applied by the designer before deployment of the AIA with the express intention of influencing a user's TRB/trust (whether or not the means for doing so conforms to a formal trust concept). As such, it is possible to design assuring properties into the system a priori. It is likewise possible to design assuring behaviors into an AIA. From the literature, there are some high-level ideas that surround the calculation of assurances: 

    \begin{itemize}
        \item Quantifying Uncertainty
        \item Reducing Complexity
        \item Assurance by Design
        \item Planning Strategies of Assurance
    \end{itemize}

    \paragraph{Quantifying Uncertainty} Being able to quantify the different kinds of uncertainty in the AIA is necessary before attempting to express that uncertainty to a human user. There are several different kinds of uncertainty that might be considered such as uncertainty in sensing, uncertainty in planning, and uncertainty in locomotion. The general idea is that a model or method needs to be incorporated in the AIA that will represent the different kinds of uncertainty to the human user in some way. A human user could use such information to inform their trust in the `situational normality', `competence', and/or `predictability' of the AIA. 

    In the surveyed literature we have seen the following main approaches to do this. In some cases uncertainty is already represented intrinsically by the algorithms and/or models being used in the AIA. Some use the built-in statistical representations of transitions and observations as the basis of quantifying uncertainty in a POMDP robot. This is an approach that has straightforward analogs in systems that use algorithms and/or models that inherently consider uncertainty.

    Models and methods that intrinsically contain or represent uncertainty are frequently available. However, even when that it is the case -- such as with POMDPs featuring transition probabilities and observation probabilities -- there are types of uncertainties that may still not be considered. For example, there are further metrics of uncertainty beyond those intrinsically captured by a planner \cite{Aitken2016-fb,Kuter2012-bv}. Using the UGV road-network problem as an example, if the UGV calculates a distribution over possible outcomes, how favorable is that distribution? Or, given a certain road-network what kind of accuracy can be expected from the POMDP solver that the UGV is equipped with? Generally these considerations might be described as addressing `uncertainties in application', which are those uncertainties that arise when trying to apply certain algorithms and models in various environments.

    Perhaps the most obvious (but not necessarily simple) approach is to quantify the uncertainty of a classifier, i.e. its probability of error in making a decision (regression methods have analogous approaches). Some have approached this problem by using a GP model for classification \cite{Gurau2016-hs}. In this way they, in essence, construct a model from empirical training data and used that model to quantify uncertainty in different test scenarios. In contrast, we might quantify the uncertainty of a classifier based solely on the input itself \cite{Zhang2014-he}. Of course these methods share common drawbacks of being solely supervised learning approaches, however even in the unsupervised domain such as reinforcement learning, methods for avoiding highly uncertain (un-safe) rewards use some form of external expert knowledge (see \cite{Garcia2015-rs, Lipton2016-dq}). 

    Uncertainty can be easier to assess if some kind of oracle, or reference is available for comparison. Quantifying the similarity between the empirical experience and the available reference can be a measure of uncertainty \cite{Kaipa2015-hy}. Of course this approach loses its appeal when a truth model isn't available. This shouldn't detract from the intent of finding some kind of reference (truth or otherwise) in which the reasoning, sensing, and other processes of an AIA can be compared to evaluate uncertainty. The evaluation of statistical models involves a very similar concept: given a statistical model as a reference does current empirical experience support or detract from the hypothesis that the model is still valid \cite{Laskey1995-jp,Ghosh2016-dl}? These approaches can quantify the degree to which the statistical models are still true, and this measurement can be used as an indication of uncertainty.

    Generally, the capability of quantifying uncertainty enables an AIA to be able to express assurances related to the `situational normality', `competence', and `predictability' of the system in a given situation. One might imagine that, in the UGV road-network problem, the UGV expressing high uncertainty in its plan would influence the competence component of the user's trust. Conversely, if an uncertainty measure is not available the user might take this as an implicit assurance that the AIA is perfectly confident, or based on the user's experience they might conclude that since all AIA plans have been flawed in the past, the plan of this AIA must be flawed as well.

    \paragraph{Reducing Complexity} Many researchers have attempted to remove complexity from the models and logic of the AIA to make the methods more interpretable (or comprehensible, or explainable, \ldots) to a human user. As with quantifying uncertainty, making an AIA more interpretable can also inform a user's trust regarding the `situational normality', `competence', and `predictability' of it. Of course this presupposes that many of the methods used by AIAs are `complex' by some measure, we claim that the fact that experts are required to understand some of the methods (and even then it may not be totally possible) proves this supposition. Complexity only exists in the presence of some reference frame, which is the designer's in this case. Generally complexity is said to increase with the number of variables, steps of reasoning, the size of data, etcetera.

    In practice, reduction of complexity has been addressed by approaches as simple as finding summary statistics, or calculating averages \cite{Muir1994-ow,Muir1996-gt}. This can also be accomplished by computing heuristic measures which reduce many complex goals into more manageable pieces\cite{Aitken2016-fb}. Creating variable fidelity models is another way by which complexity can be reduced (and increased when necessary). Attempts should be made to make/discover/learn models with scalable interpretability based on given criteria like required depth of understanding, level of expertise, and time to gain understanding.
    
    One might focus on creating and using models that are inherently more interpretable to humans (i.e. \cite{Caruana2015-za} and others). This could include constraining the feature space to be more simple, reducing dimensionality, learning more understandable features, and theoretically founding the models (i.e. interpretable science).
    
    While it is possible that there are inherently interpretable models that can be designed that can compete with other non-interpretable models, we believe that this is not the best long-term approach to reducing complexity; this is primarily due to the lack of scalability for engineers and scientists to frequently design new algorithms. Investigating methods that generate explanations from non-interpretable models is a more promising direction. The main reason is that the idea of interpretable models is not well-defined and, in reality, doesn't exist as a single tangible goal. Instead there is a continuum of interpretability that is based on the complexity of the problem, the time required for a user to interpret (i.e. a few seconds or months of study), the expertise of the user, and others. Investigating the generation of interpretations and explanations that are user specific and model agnostic would be the best of both worlds. These ideas are much more aligned with the efforts of \cite{Ruping2006-xj} and others who seek to use models with scalable resolution and accuracy.

    \paragraph{Assurance by Design} No matter how much engineers like to think about automating everything, realistically a human will need to be involved at some level of assurance design for the foreseeable future, if only because the main pursuit of human-AIA trust directly involves a human. The above two approaches alone (quantifying uncertainty, and reducing complexity) can largely use existing methods, however some researchers directly engineered their methods and models in the AIA to be more meaningful to humans.

    One approach is putting a human in the learning process, which essentially modifies the objective function of the learning algorithm \cite{Freitas2006-qo}. In essence the objective function would then be based on a large set of human preferences (and biases). This kind of approach is promising, in that it can be used to encode many human qualities that cannot be easily quantified, or even explained. However, there are trade-offs that can be undesirable in many situations as well. We often use designed, objective, learning algorithms \emph{to avoid} human biases. It is interesting to note that using a human in the loop can offer more interpretability to the result of a learning process, while at the same time making the learning process itself less procedural.

    One source of discord between what humans expect and what AIAs actually do (i.e. less predictable AIAs) is a poorly designed objective function \cite{Amodei2016-xi}. This might be referred to as myopic objectives, when an AIA focuses on a specific objective to the extent that a human can no longer relate to the objective of the AIA (and will be correspondingly surprised by its actions). This suggests that significant time may be required to design objectives that align with those of humans, this alignment will automatically make the AIA more predictable, and competent in the user's eyes.

    It is also possible to modify standard learning approaches (like the ones discussed in the previous section), in such a way as to make the methods inherently more assuring to a human. For example one might restructure a neural net architecture, or a decision tree with the sole purpose of making it more interpretable \cite{Choi2016-by,Abdollahi2016-vn,Jovanovic2016-gw}. It is difficult however, because it is not always clear how this can/should be done.

    \paragraph{Planning Explicit Assurances} Planning assurances is critical when trying to attain desired TRBs from a human user. In this context when we say `planning explicit assurances' we mean formulating a plan for the expression of assurances over time, with the goal of more effectively and appropriately expressing assurances to a human user. Having said this we recognize that planning is not a capability available to all AIAs. In cases where AIAs don't have the ability to plan, they may be designed beforehand with some kind of static plan of assurance. Otherwise, more advanced AIAs might take into account TRBs to plan an assurance strategy to assist the human to use it appropriately.

    When planning assurances the AIA must be able to account for limitations of users, and its own limitations in expressing assurances. For example a user may not be able to understand the necessary information needed to use the AIA more appropriately. Also, the AIA may need to take a long-term strategy to teach the user, as opposed to only presenting the user with canned, static, assurances. Some of the important user considerations will be discussed further in Section~\ref{sec:express_assurances}.
    
    One must ask whether the human user can correctly process the information received. This is perhaps most easily illustrated by considering a non-expert user who cannot understand highly technical assurances regarding the AIA. However, less trivial manifestations may be troubling, such as the existence of bias in the perception of assurances. This will be addressed further in Section~\ref{sec:express_assurances}.

    This topic is nearly unexplored in the context of human-AIA trust relationships. However, there are several fairly new programs that are interested in this question (i.e. explainable artificial intelligence (XAI) \cite{Gunning2016-kb}, and assured autonomy \cite{Neema2017-bb}). Assuming an AIA can give assurances there are important questions like: what is the best way to present them? How can they be adapted for different kinds of users? How can the AIA teach or tutor the human over time? This is a large gap in the current assurances landscape, answers to these questions are critical to designing more robust and effective assurances.

\subsection{Expression and Perception of Assurances} \label{sec:express_assurances}
    Expression and Perception of assurances have been combined in this section because they share several critical aspects. The key points to be considered in design of assurances are:
    
    \begin{itemize}
        \item Mediums
        \item Methods
        \item Efficacy
    \end{itemize}
    
    For an explicit assurance the medium, and method of expression must be selected taking into consideration the limitations of the AIA. Here medium denotes the means by which an assurances is expressed, this could be through any of the senses by which humans perceive, such as sight, sound, touch, smell, and taste. The method of assurance is the way by which the assurance is expressed. An example may help: a plot may be conveyed through sight in the typical way, or through spoken language (for example when communicating to a blind person); in this case the plot is the method, and sight or sound are the different mediums through which it can be communicated. An AIA might be limited in methods of expression because it does not have a display, or a speaker. In that situation how is the user supposed to receive an assurance?

    A designer must also consider whether a human can perceive the assurances being given. If so, to what extent is the information from the assurance transfered, or how efficacious was the assurance? A few examples include: an AIA giving an auditory assurance in a noisy room and the user not hearing it (such as an alert bell in a factory where the workers use ear-plugs), or an AIA attempting to display an assurance to a user that has obstructed vision. If an assurance is not expressed, or not perceived by the user, it is useless and has no effect. For example, an AIA may have the ability to store data about its performance, and compute a statistic regarding its reliability, but if it cannot successfully express (or communicate) that information in some way, the information is useless.

    A user will \emph{always} have some kind of TRB towards an AIA (if only to choose to ignore the AIA). In the absence of explicit assurances the human will instead use implicit assurances to inform their TRBs. However, the general human user will not have knowledge regarding which assurances are implicit or explicit -- humans participating in research from Quadrants I and II were not aware which assurances were designed by the researchers and which weren't. Recall from Section~\ref{sec:assurances} that to a user all assurances are the same; that is to say that any property or behavior of an AIA that affects trust is an assurance to a user, and it doesn't matter whether the assurance was designed for that purpose or not (is explicit or implicit).

    \paragraph{Mediums:} One might use sight to express an assurance. For example an AIA might give visual feedback, or display different performance characteristics \cite{Chadalavada2015-wx,Muir1996-gt}. This can also be accomplished by using natural language \cite{Wang2016-id} -- it is also a simple matter to convert written natural language output to spoken natural language now. Other typical mediums are blinking lights, colored boxes, ringing bells, buzzers, recorded voice alerts and others.
    
    The other senses (touch, smell, and taste) are not well explored in literature related to human-AIA trust. Generally, any human sense could be used as a medium. Besides sight, and sound, one of of tactile feedback has been used extensively in robotics where it is called `haptic feedback' (where the user receives mechanical feedback through the robot controls). This medium is use to create a more immersive user interface in robotics, to help users feel more connected to the robot. One can imagine smell and taste having an obvious application in the assurances of a cooking robot, other applications certainly exist as well and are open to further research.

    \paragraph{Methods:} One of the main methods by which to express an assurance is by displaying a value, such as a flow-rate \cite{Muir1996-gt}. While this may sound banal it actually involves some nuanced points. The interesting part is that a value such as a flow-rate actually conveys no assurance to a human user without the human user then creating a corresponding mental model of the trustworthiness of an AIA capability. The user's trust dimensions (`competence', 'predictability', etcetera) are then affected by this perception. This approach (also used by \cite{Wickens1999-la,Sheridan1984-kx,Hutchins2015-if} and others) is effective, but relies heavily on the assumption that the user will create a model that is `good enough' out of sequentially presenting certain information.

    Similarly one might train a user to recognize signs of failure in different interactions with an AIA \cite{Freedy2007-sg,Desai2012-rc,Salem2015-md}. Still this approach relies on human users to make models that are `good enough' in order to correctly decide how to appropriately use the AIA. The main drawback of this work, and of that above is the blind reliance on users being able to make correct statistical models (of things like reliability) from noisy observations. A more ideal approach would be to design assurances that remove chances for misinterpretation because of inconsistent human models.

    More direct methods of expressing assurances include displaying the intended movements through visual projection of a planned path \cite{Chadalavada2015-wx} -- this is subtly, but significantly different from making the user infer the intended intention. Analogously, natural language expressions (written or otherwise) attempt a more active method of expressing an assurance (such as \cite{Wang2016-id}). One might also display plans and logic in different formats such as tables or trees, bar charts  \cite{Van_Belle2013-ph, Huysmans2011-th, Hutchins2015-if}. These approaches attempt to remove some uncertainty regarding the human's ability to create the correct model. As humans are fond of saying ``You can't assume that I can read your mind!'', in essence more passive expressions from AIAs are relying on humans to read AIA's `minds' (we can't even do that with other humans).

    Any of these methods can be more or less effective based on the task, or situation in which they are used. How should uncertainly be displayed (i.e. as a distribution, summary statistics, fractions or decimals)?  Unsurprisingly we find that the answer is `it depends' \cite{Chen2014-dk,Wallace2001-fm,Kuhn1997-qc,Lacave2002-cu}. Things such as the experience of the user, or the nature of the information being displayed affect the user's ability to interpret the assurance. In the absence of that information the best that can be done is to select the method that will work for the largest group of typical users of the AIA. A sufficiently advanced AIA might also be able to learn the best methods to communicate to different users on a one-on-one basis.

    It is generally presumed that making something more human-like will make an AIA more trustworthy. An algorithm may be human-like when it represents knowledge in a way that a human would understand, or executes logic in a way that a human can follow. A robot that is humanoid becomes more human-like in appearance \cite{Bainbridge2011-pl}, a system that uses natural language becomes more human-like in communication \cite{Lacave2002-cu}. The human-AIA trust relationship depends on assurances that, in essence, are conversions from AIA capabilities to human-perceptible behaviors and properties.  Assurances are the method of communication upon which humans can learn to trust AIAs. Because of this it is expected that at some point all assurances have to at least be made human-understandable in some way, otherwise the AIA is essentially speaking a different language, and assurances cannot be understood.

    Making something `human-like' doesn't always increase it's trustworthiness. In \cite{Dragan2013-wd} the AIA is made more trustworthy by making the robot motions more human-like, whereas in \cite{Wu2016-ei} making the AIA more human-like resulted in a decrease of trustworthiness. In this case the difference came from the type of task, in the first case the robot was physically working in proximity to a human, in the other case the user was playing a competitive game against the AIA. It has been observed that humans trust more `human-like' AIAs in more human-like ways \citet{Tripp2011-rx}. Perhaps `human-like' applies to how difficult it is to relate to the AIA. Following on this idea the benefits or drawbacks of human-like characteristics are influenced by a user's general impressions and feelings of how trustworthy humans are in similar situations. This would also involve aspects of psychology, sociology, and is very difficult to control and account for. This is an open research question that is important to answer in order to design better assurances.

    It is worth considering, in more detail, what implications the existence of implicit and explicit assurances has on the designer. It is unrealistic to think that a designer can take all possible assurances into account, and thus will need to focus efforts on some of the most important. The foremost consideration is that an analysis of the interaction between the human and user needs to be made in order to identify the critical assurances for a given scenario. For example, in the road network problem, an analysis might find that the most critical assurances are about the competence of the UGV's planner. In this case the designer must take time to design an assurance that is directed at the user's perception of the AIA's competence -- let's call this a planning-competence assurance.

    One difficulty arises from this approach is that there doesn't seem to be a way to determine what passive assurances might drown out active assurances. Following from the example above, the designer may have built an excellent planning-competence assurance, but failed to consider the effect of how the UGV appears -- it may be old, have loose panels, and rust holes. Generally, designers overlook implicit assurances (i.e. do not consider them explicitly in design) because they assume that they will have no effect (i.e. why does it matter if there are rust-holes if the UGV works?). This can stem from either: 1) ignorance of human-AIA trust dynamics, or 2) lack of identifying which assurances are most important to a human user.

    While it might be nice, it seems unreasonable, inefficient, and unwise to attempt a study of \emph{every possible} assurance from an AIA to a human and then select the most important. Perhaps one way a designer might try to identify which assurances are important is to perform human studies where feedback about which characteristics of the AIA most affected the trust of the user. An approach like this would help to point out if explicit assurances are being noticed, and if there are implicit assurances that are overly influential, or that overwhelm the explicitly designed assurances. With such feedback designers would have a realistic idea about whether their explicit assurances are having the desired effect. We use the UGV road-network problem to illustrate. After designing an explicit assurance (or more) the operator/UGV team could work together, afterwards the operator could rank the different behaviors/properties of the AIA affected their trust in it. In this way the critical implicit and explicit assurances will be identified. If the explicit assurance is near the top of the list of influencing assurances then it is working, if not a re-design may need to occur.

    One final point is that there are several potential sources of explicit assurance that lack appropriate expressions, and thus cannot be effectively utilized as assurances. For example, it is unclear the best way for an AIA to express that it has been validated and verified on situations similar to the current one. Similarly, what other methods exist for communicating statistical distributions besides showing a plot (only useful for 1 or 2 dimensional distributions) or showing sufficient statistics? Investigating how assurances can be expressed in effective, and efficient ways is critical to human-AIA trust relationships.

    \paragraph{Efficacy:} Some kinds of expression are very `one-dimensional' in that they only use one medium, or method. This, again, has been seen in practice by the utilization of plotting a certain value over time. Because of this, much of the research to date involves assurances that are not robust to loss in transfer, meaning that the approaches count everything on a specific medium and method to work, and if not the whole assurance is rendered useless. Hence, exploring ways in which assurances can be robustly communicated is a clear opportunity for those trying to design assurances. This is akin to a human speaking with their voice, making facial expressions, and gestures with their hands as well; simultaneously utilizing several mediums/methods helps to ensure an assurance will be effective. Of course, repeating the same message over a thousand times is wasteful, and so enters the idea of efficiency in expression.

    Perhaps less obvious is a situation in which the user has to supplement an incomplete assurance. A user can supposedly create a mental model of the trustworthiness of an AIA capability -- based on repeated observations over time. Creating this mental model takes time, and the model is prone to cognitive biases. In this case the assurance is communicated slowly and indirectly. Generally, a highly effective assurance would have precise information communicated in a way that is easy for the user to perceive, with little loss. Whereas, an inefficient, and ineffective assurance may be more vague, wasteful (i.e. repeating the same thing a thousand times), and susceptible to loss in communication. The solutions to efficacy lay in selecting appropriate methods, and mediums for expression of the assurance, and by designing for appropriate levels of redundancy to ensure that the assurance is received.

\subsection{Observing Effects of Assurances} \label{sec:measuring_effects}

    Since the purpose of assurances is to affect TRBs, it is important to be able to quantify those effect. There are two different instances in which this is useful: 1) when the designer needs an understanding of how effective the assurances are; and 2) When the  AIA needs to observe the efficacy of it's own assurances.
    
    To our knowledge there has not been any work regarding the second situation that enables an AIA to observe responses to assurances and then adapt behaviors appropriately (at least not in the trust cycle setting), however this is arguably the ultimate goal so that AIAs can themselves modify assurances to meet different needs. What does this mean practically? Theoretically, any method that is made for the designer to measure the effects of assurances could also be deployed into an AIA. The surveyed literature gives some insights into how that has been done to date.
   
    When it comes time to measure the effect of assurances on a human's trust there are two main approaches:
    
    \begin{itemize}
        \item Gather self-reported changes in trust from human users
        \item Measure changes in user's TRBs
    \end{itemize}
    
    \paragraph{Self-Reported Changes} Gathering self-reported changes in trust involves asking questions like: `how trustworthy do you feel the system is?'; or `to what extent do you find the system to be interpretable?' \cite{Mcknight2011-gv,Muir1996-gt,Wickens1999-la,Salem2015-md,Kaniarasu2013-ho}. These kinds of questions can be useful in verifying whether the assurances are having the expected effect. It is not unreasonable to imagine that an AIA might be equipped with a method by which it can ask the user questions about their trust, process those responses, and modify assurances appropriately.
    
    However, sometimes changes is self-reported trust do not result in changes in TRBs \cite{Dzindolet2003-ts}. From the AIAs perspective this means that --- unless the object of the assurances is to make the person's level of self-reported trust change --- the assurances are not providing any benefit. As previously discussed, the goal of assurances is to elicit appropriate TRBs from the human user. From this perspective, measuring changes in TRBs is the more direct, and objective, approach to measure the effect of assurances.

    Self-reports are the most useful when trying to understand the true effects of an assurance. Does a certain assurance, assumed to affect `situational normality', actually do that? There is space for quite a lot of research in this realm. Does displaying a specific plot actually convey information about `predictability'? This information can be used to inform the selection of the methods of assurance.

    \paragraph{Measuring Changes in TRBs} Generally researchers in the field have measured, in some way, how frequently the AIA was able to run in autonomous mode before being turned off \cite{Freedy2007-sg,Desai2012-rc}. Other researchers calculated whether the user was willing to cooperate with the AIA or not \cite{Salem2015-md,Wu2016-ei,Bainbridge2011-pl}. A better defined metric is: the likelihood of appropriate use of a certain capability by the user; albeit more difficult to formally define/calculate in different situations. As a concrete example, in the UGV road-network problem there isn't really an option to `turn off' the UGV. Instead the remote operator can make decisions such as accepting a plan designed by the UGV. In this situation the effect of assurances might be measured by how likely the operator is to accept a generated plan instead of overriding it (recall that the goal may not be to have the generated plan accepted 100\% of the time, rather that it be accepted with respect to how appropriate it is in a given situation).

    In practical application (such as in the UGV road-network problem), the user, and the human-AIA team, care more about whether TRBs are appropriate or not. It doesn't help if an assurance helps the user feel that the AIA is more competent, if the user doesn't treat the AIA any differently than before the assurance. This assumes that it is possible for appropriate TRBs to be measured in the first place. For example, if appropriate behavior is for the user to verify a sensor reading, can the AIA perceive that happening? In that situation perhaps the easiest approach would be to ask the user, but what if the user is dishonest? Is there a way to verify the user behavior is actually appropriate? This is something that has gained notoriety with the current generation of autonomous cars, where the car is capable of autonomous operation, but the user still needs to sit in the driver's seat and be attentive just in case the vehicle cannot perform correctly. These are some arguments for the importance of designing methods for perceiving appropriate (and inappropriate) TRBs when designing assurances. 
%
%
%

\section{Conclusions}\label{sec:conclusions}
    Now, more than ever, there is a great need for humans to be able to trust the AIAs that we are creating. Assurances are the method by which AIAs can encourage humans to trust them appropriately, and to then use them appropriately. We have presented here a definition, case for, and survey of assurances in the context of human-AIA trust relationships.
    
    This survey was performed, to some extent, from a standpoint of designing an unmanned ground vehicle that is working in concert with a human supervisor. However, the theoretical framework, and classification of assurances is meant to be general in order to apply to a broad range of AIAs. One of the main motivations of this survey was the insight that there is an extremely large community of researchers working on human-AIA assurances (perhaps unknowingly). It is important to recognize this so that we can start to organize our efforts and begin methodically answering the open questions of this important field. Arguably, the ultimate goal is to develop a sufficient set of assurances that can be located in Quadrant III (i.e. those that have been designed and experimentally verified), of course this requires cooperation among the community.

    The most surprising insight from compiling this survey was the absence of a detailed definition and classification of assurances. Assurances have been, by far, the most ignored component of human-AIA trust relationships. There have been many researchers who have recognized the existence of assurances (usually by other names), but there has been no detailed definition until now. We have drawn from multiple bodies of research in order to `fill in' the details of the human-AIA trust cycle (Figure~\ref{fig:SimpleTrust_one_way}) (a novel contribution in itself). This led to our main contribution: to formally define and classify assurances within the human-AIA trust cycle. We have introduced the idea that assurances must stem from different AIA capabilities, that they can be implicit or explicit, that the way in which they are expressed to a user can be just as important as how they are calculated/designed. We have also highlighted that methods for measuring the effects of assurances must also exist. And, we have shown how each of these fits into the larger trust cycle.
    
    There is a fairly large body of research that is focused, in some way, on influencing trust in human-AIA relationships. However, there is a larger portion of research that deals with techniques that would be useful in designing assurances but that, to date, has not directly or knowingly considered affecting human-AIA trust through assurances as a formal design goal. Research from these areas (such as V\&V, active learning, and safety) should provide a rich collection of methods to be studied and formally applied to human-AIA trust relationships.

    While the basic definition of assurances (i.e. feedback to user trust, in the human-AIA trust cycle) is simple from a theoretical standpoint, the exercise of gathering related literature helped to illuminate some important considerations and details regarding the design of assurances. In Section~\ref{sec:synthesis} we present Figure~\ref{fig:refined_assurances} which is an original synthesis of the surveyed literature. We show that designers must be able to design ways for an AIA to calculate, design, and plan explicit assurances. Designers must also account for the expression and perception of assurances, this involves considering how effective a given method/medium might be in conveying an assurance to a user. Finally, the whole purpose of assurances is lost if there is no way to measure/quantify the effects of the assurances; effort must be spent in creating appropriate metrics and experiments by which to do this.

    A sobering reminder is that there is not a single assurance that will perform the best in all situations. It is almost certain that given time highly specialized assurances can be designed for many situations. Even so we warn that, for the research and design of assurances to be sustainable in the current environment of fast-paced development of new technology, it is important to consider approaches that are as general as possible in order to be more easily used with newly developed methods for implementing AIA capabilities.

    The treatment of assurances in this survey are based, in part, on a model of trust. For completeness it is important to mention distrust. As reviewed and discussed by \citet{Lewicki1998-ox}, and formalized in \cite{McKnight2001-hm,McKnight2001-gz}. Low trust is not the same as distrust, neither is low distrust the same as trust. \citet{McKnight2001-gz} suggest that ``the emotional intensity of distrust distinguishes it from trust'', and they explain that distrust comes from emotions like: wariness, caution, and fear. Whereas, trust stems from emotions like: hope, safety, and confidence. Trust and distrust are orthogonal elements that define a person's TRB towards a trustee. In this survey distrust was not considered, however it must be made clear that any \emph{complete} treatment of trust relationships, and for our purposes, designed assurances, must consider the dimensions of distrust as well as those of trust. For now, this investigation is left as an avenue for future research.

    We hope that researchers can begin to reach across perceived lines in order to find more tools to appropriately design and test assurances. We hope that the material is Section~\ref{sec:background} will provide a common foundation on which researchers from all quadrants can build on in order to unify research efforts. More specifically, we hope that those in the fields of CS, ML, and AI can begin to use the principles outlined in this survey to help guide their search for more `interpretable', `explainable', and `comprehensible' systems. It is important for them to understand the existence of a trust model, and the human-AIA trust cycle. If they consider these points, they will be able to identify better methods, and design more effective assurances.
    
    Likewise, those who formally consider trust (i.e. researchers in HRI, e-commerce, UI) should now be able to identify more methods and approaches for designing assurances, so that they can perform experiments to validate them. They will also have a better idea of what kinds of experiments have been performed, and possible new areas to investigate.

    We have identified many opportunities for further research in how AIAs can influence human trust through assurances. The framework found herein will help other researchers to see the field from a larger perspective, to classify the type of research they are performing, and help them to consider the greater implications of their work.

\bibliographystyle{ACM-Reference-Format}
\bibliography{References}
\end{document}